\documentclass[12pt]{iopart}

\usepackage{graphicx}
\usepackage{wrapfig}
\usepackage{subcaption}
\usepackage[margin=0.5cm]{caption}
\usepackage{amssymb, bm}
\usepackage{color}

\usepackage{ulem}

\def\ns#1{neutron star#1 (NS#1)\gdef\ns{NS}}
\def\gw#1{gravitational wave#1 (GW#1)\gdef\gw{GW}}
\def\bh#1{black hole#1 (BH#1)\gdef\bh{BH}}
\def\bbh#1{binary black hole#1  (BBH#1)\gdef\bbh{BBH}}
\def\bns#1{binary neutron star#1 (BNS#1)\gdef\bns{BHS}}
\def\bhns#1{black hole - neutron star#1 (BHNS#1)\gdef\bhns{BHNS}}
\def\nr#1{numerical relativity#1 (NR#1)\gdef\nr{NR}}
\def\Nr#1{Numerical relativity#1 (NR#1)\gdef\Nr{NR}}
\def\ligo#1{Laser Interferometer Gravitational-wave Observatory#1 (LIGO#1)\gdef\ligo{LIGO}}

\begin{document}

\title{Black Hole - Neutron Star Binary Mergers: The Imprint of Tidal Deformations and Debris}

\author{Bhavesh Khamesra$^{a}$,
		Miguel Gracia-Linares$^{b}$,
        Pablo Laguna$^{b}$
}
\address
{$^a$Center for Relativistic Astrophysics and School of Physics, Georgia Institute of Technology, Atlanta, GA 30332, U.S.A.\\
$^b$Center for Gravitational Physics, Department of Physics, The University of Texas at Austin, Austin, TX 78712, U.S.A.
}

\ead{ bhaveshkhamesra@gatech.edu, mgracia@austin.utexas.edu, pablo.laguna@austin.utexas.edu}

\begin{abstract}
The increase in the sensitivity of gravitational wave interferometers will bring additional detections of binary black hole and double neutron star mergers. It will also very likely add many merger events of black hole - neutron star binaries. Distinguishing mixed binaries from binary black holes mergers for high mass ratios could be challenging because in this situation the neutron star coalesces with the black hole without experiencing significant disruption. To investigate the transition of mixed binary mergers into those behaving more like binary black hole coalescences, we present results from merger simulations for different mass ratios. We show how the degree of deformation and disruption of the neutron star impacts the inspiral and merger dynamics, the properties of the final black hole, the accretion disk formed from the circularization of the tidal debris, the gravitational waves, and the strain spectrum and mismatches. The results also show the effectiveness of the initial data method that generalizes the Bowen-York initial data for black hole punctures to the case of binaries with neutron star companions.
\end{abstract}

\section{Introduction:}

The Gravitational-Wave Catalogue by LIGO and Virgo has been recently updated to bring the total number of detections to 50 \cite{Abbott:2020niy}, with 46 of the events confirmed \bbh{} mergers and two double \ns{s} mergers (GW170817~\cite{TheLIGOScientific:2017qsa} and GW190425~\cite{Abbott:2020uma}). Although not fully confirmed, the remaining two detections (GW190814 \cite{Abbott:2020khf} and GW190426\_152155 \cite{Abbott:2020niy}) suggest that the \gw{s} detected were produced from mergers of \bhns{} binaries. As LIGO and Virgo reach design sensitivity, we will have more \gw{} detections from \bhns{} binaries. Characterising these events calls for numerical simulations that are not only more accurate but that include the relevant micro-physics.

Numerical studies of \bhns{s} have considered different aspects of the merger. Some have focused on the formation of the accretion disk from the tidal debris as well as the relativistic jets emanating from the remnant \bh{}. Specifically, the studies have investigated how the accretion disk, ejecta and the jets depend on the mass ratio of the binary~\cite{Hayashi:2020zmn,Foucart:2019bxj,Shibata:2009cn, Etienne:2007jg, Foucart:2011mz}, and the spin magnitude and orientation of the \bh{} \cite{Foucart:2011mz, Foucart:2010eq, Foucart:2012vn, Foucart:2016vxd, Kawaguchi:2015bwa, Etienne:2008re,  Lovelace:2013vma, Kyutoku:2011vz,Ruiz:2018wah}. The studies have also looked at the impact of the characteristics of the \ns{,} such as  its spin \cite{Ruiz:2020elr}, magnetic field \cite{Etienne:2011ea,Ruiz:2018wah, Paschalidis:2014qra, PhysRevD.86.084026} and  equation of state \cite{Foucart:2012vn, Kawaguchi:2015bwa, Foucart:2014nda, Duez:2009yy,Brege:2018kii, Kyutoku:2011vz,Kyutoku:2010zd,Kyutoku:2015gda,Kyutoku:2017voj}. For low mass ratio systems with highly spinning \bh{} and/or lower compactness of the \ns{}, the final \bh{} is typically surrounded with massive accretion disks with densities $\geq 10^{12}~\textrm{g/cm}^3$ \cite{Shibata:2011jka}. On the other hand, for systems with high mass ratio and low \bh{} spin, the \ns{} barely suffers any disruption before reaching ISCO and can be swallowed almost completely by the \bh{} hardly leaving any trace of matter to generate detectable electromagnetic signatures. In the absence of any significant disruption, the \bhns{} systems behave as a \bbh{}, with almost identical \gw{} signatures \cite{Foucart:2013psa}.

The work in this paper has two main objectives. One is to test the effectiveness of the initial data method introduced in Ref.~\cite{Clark:2016ppe}. The method generalizes the Bowen-York~\cite{PhysRevD.21.2047} approach for initial data with \bh{s} modeled as punctures to the case of \ns{s}. The second is to provide further insights on the transition of a \bhns{} into a \bbh{-}like behavior as the effects from the disruption of the \ns{} change with the mass ratio of the binary. Our results show that for low mass ratio cases, a considerable amount of energy and angular momenta, that otherwise would have been radiated in \gw{s}, gets trapped in the accretion disk and redistributed as the \bh{} accretes the material. The tidal debris also affects the ringing of the final \bh{} when compared with the \bbh{} case. For all the cases considered, the \bhns{} binary merges earlier than the corresponding \bbh{}. This is due to the tidal deformation that the \ns{} experiences. The deformation introduces a correction to the potential that increases the orbital velocity and thus the emission of \gw{s}~\cite{Shibata:2011jka}. Our results have limitations since we model the \ns{} as a polytrope and do not include magnetic fields or neutrino transport. At the same time, we demonstrate that the initial data method has promising feature, such as simplicity of implementation and generalization to realistic equations of state.

The paper is organized as follows: Section \ref{sec:init_data} provides a summary of the initial data method developed in \cite{Clark:2016ppe}. Section \ref{sec:initparams_nrsetup} details the parameters of the initial \bhns{} and \bbh{} configurations. The section also includes the setup of the numerical simulations and convergence tests.  Results are presented in Section \ref{sec:results} organized by i) inspiral and merger dynamics, ii) the final \bh{}, iii) tidal debris, iv) \gw{s}, and v) spectrum and mismatches. Conclusions are given in Section~\ref{sec:conclusions}. We use geometrical units in which $G=c=1$ and express all dimensions in terms of $M$, the total initial mass of the binary system. When necessary, we will also use physical units (SI units). Indices with Latin letters from the beginning of the alphabet denote space-time dimensions and from the middle of the alphabet spacial dimensions.

\section{Initial Data}
\label{sec:init_data}

We will briefly review the approach we introduced in Ref.~\cite{Clark:2016ppe} to construct initial data for binaries with \ns{} companions. Under the 3+1 decomposition of the Einstein equations, initial data consist of the spatial metric $\gamma_{ij}$ of the constant time initial hypersurface, the extrinsic curvature $K_{ij}$ in this hypersurface, and the projections
\begin{eqnarray}
\label{eq:hmdensity_pf}
\rho_H &\equiv n^a n^b T_{ab}\\ 
\label{eq:momdensity_pf}
S^i &\equiv -\gamma^{ij} n^b T_{jb}
\end{eqnarray}
of the stress-energy tensor $T_{ab}$, with $n^a$ the unit time-like normal to the hypersurface. For the present work we will only consider perfect fluids. Thus,
\begin{equation}
\label{eq:stressenergy}
T_{ab} = (\rho + p)u_a u_b + p g_{ab},
\end{equation}
with $\rho$ the energy density, $p$ the pressure, $u^a$ the four velocity of the fluid, and $g_{ab} = \gamma_{ab}-n_an_b$ the space-time metric. With this form for $T_{ab}$,
\begin{eqnarray}
\label{eq:hmdensity_pf}
\rho_H &=&  (\rho + p)\,W^2 - p\\
\label{eq:momdensity_pf}
S^i &=&  (\rho + p) \,W\,u^i,
\end{eqnarray}
where $W = -n_au^a$ is the Lorentz factor, which can be rewritten as
\begin{equation}
        W^2 = \frac{1}{2}\left(1+\sqrt{1 + \frac{4 S_iS^i}{(\rho + p)^2}}\right).
\end{equation}
The initial data $\lbrace \gamma_{ij}, K_{ij}, \rho_H, S^i \rbrace$ must satisfy the constraints
\begin{eqnarray}
\label{eq:ham}
    R + K^2 - K_{ij} K^{ij} &= 16 \pi \rho_H\\
\label{eq:mom}
    \nabla_j \left(K^{ij} - \gamma^{ij}K \right) &= 8 \pi S^i,
\end{eqnarray}
namely the Hamiltonian and momentum constraints, respectively. Here, $R$ is the Ricci scalar, and $\nabla_j$ is the covariant derivative associated with $\gamma_{ij}$.

We solve Eqs. (\ref{eq:ham}) and (\ref{eq:mom}) following the conformal-transverse-traceless (CTT) approach pioneered by Lichnerowicz~\cite{
baumgarte_shapiro_2010}, York and collaborators~\cite{Smarr:1979ofa}. The central idea of this approach is to apply the following transformations to isolate the four quantities obtained by solving the constraints:
\begin{eqnarray}
    \gamma_{ij} &=& \psi^4 \tilde{\gamma}_{ij}\\
    K_{ij} &=& A_{ij} + \frac{1}{3}\gamma_{ij}K \\
    A^{ij} &=&\psi^{-10} \left(\tilde{A}_{TT}^{ij} + \tilde{A}_L^{ij}\right),\\
     \widetilde\nabla_j\tilde{A}^{ij}_{TT} &=& 0\\
\tilde{A}_L^{ij} &=& 2\widetilde\nabla^{(i} \mathcal{V}^{j)}  - \frac{2}{3}\tilde{\gamma}^{ij}\, \widetilde\nabla_k \mathcal{V}^k \\
    \label{eq:ch2_hd_ct}
    \tilde{\rho}_H &=& \rho_H \psi^8\\
    \label{eq:ch2_md_ct}
     \tilde{S}^i &=& S^i \psi^{10},
\end{eqnarray}
where $\psi$ is the conformal factor. The last two transformations imply that $\tilde{\rho} = \rho\, \psi^8$, $\tilde{p} = p\, \psi^8$, $\tilde{u}^i = u^i\psi^2$ and $\widetilde W = W$.

We also adopt the common choices of conformal flatness ($\tilde{\gamma}_{ij} = \eta_{ij}$), maximal slicing ($K=0$), and $\tilde{A}^{\rm TT}_{ij} = 0$. With these choices and the CTT transformations above, the Hamiltonian and momentum constraints take the following form:
\begin{eqnarray}
    \label{eq:ham2}
    \tilde{\Delta} \psi+ \frac{1}{8}\psi^{-7}\tilde{A}_{ij}\tilde{A}^{ij} &= -2\pi \psi^{-3}\tilde{\rho}_H \\
    \label{eq:mom2}
    \widetilde{\nabla}_j \tilde A^{ij} &= 8 \pi \tilde{S}^i.
\end{eqnarray}
Bowen and York \cite{PhysRevD.21.2047} found point-source solutions to the source-free momentum constraint (\ref{eq:mom2}) that can be used to represent \bh{s} with linear momentum $P^i$ and spin $J^i$. The solutions read:
\begin{eqnarray}
  \tilde A^{ij} &=& \frac{3}{2\,r^2}\left[ 2\,P^{(i} l^{j)} - ( \eta^{ij}-l^il^j)P_k l^k\right]\label{eq:KP}\\
  \tilde A^{ij} &=& \frac{6}{r^3}l^{(i}\epsilon^{j)kl}J_kl_l  \label{eq:KS}
\end{eqnarray}
where $l^i = x^i/r$, a unit radial vector.

In Ref.~\cite{Clark:2016ppe}, we followed Bowen's approach \cite{Bowen1979} to construct solutions to the momentum constraint that represent \ns{s}. The solutions assume spherically symmetric sources and are given by
\begin{eqnarray}
\tilde{A}^{ij} &=& \frac{3Q}{2r^2}\left[2P^{(i}l^{j)} - (\eta^{ij} - l^il^j)P_kl^k  \right] \nonumber\\
&+& \frac{3C}{r^4}\left[2P^{(i}l^{j)} + (\eta^{ij} - 5l^il^j)P_kl^k  \right]\label{eq:KP2}\\
\tilde{A}^{ij} &=& \frac{6\,N}{r^3}l^{(i}\epsilon^{j)kl}J_k l_l\label{eq:KS2}\,,
\end{eqnarray}
where
\begin{eqnarray}
	\label{eq:qjcdefs}
	Q &=&4\pi \int_0^r  \sigma \bar r^2 \, d\bar r \\
	C &=& \frac{2\pi}{3} \int_0^r  \sigma \bar r^4 \, d\bar r\\
	N &=& \frac{8\,\pi}{3}\int_0^r \chi\,\bar r^4\,d\bar r\,.
\end{eqnarray}
The source functions $\sigma$ and $\chi$ are radial functions with compact support $r\le R$ and are such that
\begin{eqnarray}
\label{eq:ch2_conf_momden}
        \tilde{S}^i &= P^i \sigma\\
        \tilde{S}^i &= \epsilon^{ijk} J_j x_k\chi\,.
\end{eqnarray}
Outside the sources, $Q=N=1$ and $C=0$; thus, the extrinsic curvatures (\ref{eq:KP2}) and (\ref{eq:KS2}) reduce to those of point sources, i.e. (\ref{eq:KP}) and (\ref{eq:KS}) respectively.

Since $\tilde S^i = (\tilde\rho+\tilde p) W\,\tilde u^i $, we set
\begin{eqnarray}
\sigma  &=& (\tilde\rho+\tilde p)/\mathcal K\\
\chi  &=& (\tilde\rho+\tilde p)/\mathcal N
\end{eqnarray}
with the constants ${\mathcal K}$ and ${\mathcal N}$ obtained from
\begin{eqnarray}
   {\mathcal K} &=&  4\,\pi \int_{0}^{R} (\tilde\rho+\tilde p)\,r^2\, dr \\
   {\mathcal N}&=&   \frac{8\,\pi}{3} \int_{0}^{R} (\tilde\rho+\tilde p) \,r^4\, dr\,.
\end{eqnarray}
Given these solutions for the extrinsic curvature, we solve the Hamiltonian constraint (\ref{eq:ham2}), assuming that the conformal factor has the form $\psi = 1+m_p/(2r)+u$ where $m_p$ is the bare or puncture mass of the \bh{.} To solve (\ref{eq:ham2}), we used a modified version of the \texttt{TwoPunctures} code~\cite{Ansorg:2004ds}  which handles the source $\tilde\rho_H$.

The method to construct \bhns{} initial data in Ref.~\cite{Clark:2016ppe} for a \bh{} with irreducible mass $M_h$ and \ns{} with mass $M_*$ follows similar steps to that for \bbh{s} initial data with punctures.
That is, one selects the target values for $M_*$ and the mass ratio
$q=M_h/M_*$. For \bbh{} systems, one usually chooses instead of $M_*$ the total mass $M$ of the binary.  Next, one carries out  iterations solving the Hamiltonian constraint until the target values for $q$ and $M_*$ are obtained. After each Hamiltonian constraint solve iteration, one computes $M_h$ from the irreducible mass of the \bh{.} The challenge is in finding an appropriate definition for the mass $M_*$ of the \ns{} in the binary. Options are the ADM mass $\mathcal{M}_A$ or rest mass $\mathcal{M}_0$  of the \ns{} in isolation, which in isotropic coordinates read
\begin{eqnarray}
	\mathcal M_{A} &=& 2\pi \int_0^R\rho\, \psi^5 r^2\, dr \\
    \mathcal M_0 &=&  4\,\pi \int_0^R  \rho_0\, \psi^{6} r^2\, dr
\end{eqnarray}
respectively, with $\rho_0$ the rest-mass density.
The approach we suggested in Ref.~\cite{Clark:2016ppe} is to compute the mass after each Hamiltonian constraint solve iteration  from $M_*^{(n)} = \xi^{(n-1)}\,M_0^{(n)}$ where $\xi^{(n-1)} = \mathcal M_A^{(n-1)}/\mathcal M_0^{(n-1)}$, namely the ratio of the ADM and rest mass of the star in isolation. Here
\begin{equation}
    M_0 = \int  \rho_0 W \sqrt{\gamma} d^3x =  \int  \tilde\rho_0 W \psi^{-2} d^3x
\end{equation}
is the rest mass of the \ns{} after each Hamiltonian constraint solve.
For $n=1$, $M_*^{(1)} =\mathcal M_A^{(1)}$; thus, $\xi^{(0)} = 1$. We have found that for the simulations we have considered, $\xi \approx 0.93$, with variations less than $1\,\%$ throughout the iteration procedure; thus, our method is close to those in which  the value of the rest mass of the \ns{}  is the target.

\section{Initial Parameters, Numerical Setup, and Convergence Tests}
\label{sec:initparams_nrsetup}

 We study mixed binaries with mass ratio $q = 2,\, 3$ and 5, labeled Q2, Q3, and Q5, respectively. In the present work, we will consider only non-spinning \bh{s} and \ns{s} and model the \ns{} as a polytrope, i.e. $P = \kappa\,\rho_0^\Gamma$ equation of state. In all cases, we set $M_* = 1.35\,M_\odot$, $\kappa=93.65\, M_\odot^{2}$, $\Gamma=2$, and coordinate separation $9\,M$, where $M= M_h+M_*$. The momenta $P^i$ for each compact object in the binary is obtained by solving the $3.5$ post-Newtonian equation of motion from a large separation and stopping at separation where the numerical relativity simulation begins.
Table~\ref{tab:ID} shows the initial parameters for the simulations at the end of the construction of the initial data. Our configurations closely mimic models $M20.145$, $M30.145$, and $M50.145$ in Ref.~\cite{Shibata:2009cn}, models $B$ and $A3$ in Ref.~\cite{Etienne:2007jg}, and models $A$ and $D$ in Ref.~\cite{Etienne:2008re}.

We use the \texttt{MAYA} code~\cite{2015ApJLEvans,2016PRDClark,2016CQGJani} for the simulations; the code is our local version of the \texttt{Einstein Toolkit} code~\cite{steven_r_brandt_2020_3866075}.
It solves the BSSN \cite{Shapiro1999,Shibata1995} form of the Einstein evolution equations and follows the implementation in the \texttt{Whisky} code \cite{Baiotti:2004wn,Hawke:2005zw, Baiotti:2010zf} for the hydrodynamical evolution equations. We use the Marquina solver \cite{Aloy:1999marq} to handle the Riemann problem during flux computation and the piece-wise parabolic method \cite{Colella:1984ppm} for reconstruction of primitive variables. The \bh{} apparent horizon is found using the \texttt{AHFinderDirect} code~\cite{Thornburg:2003sf}. We use two methods to track the \ns{.} One method tracks the maximum density within the star. The other tracks the star using the \texttt{VolumeIntegrals} thorn in the \texttt{Einstein Toolkit} ~\cite{steven_r_brandt_2020_3866075}. The properties of the \bh{}, mass, spins and multipole moments, are computed using the \texttt{QuasiLocalMeasures} thorn \cite{Loffler:2011ay} based on the dynamical horizons framework~\cite{2004LRR.....7...10A}. The \gw{} strain is computed from the Weyl scalar $\Psi_4$~\cite{Loffler:2011ay, Zilhao:2013hia,Reisswig:2010di}. To compute the radiated quantities, we follow the method developed in \cite{Ruiz:2007yx}. The gauge choice for the evolutions is the moving puncture gauge~\cite{Campanelli2005,Baker2006}.

\begin{table}
\begin{center}
\begin{tabular}{@{}*{10}{c}}
\br
 Case & $q$ & $M_0/M_\odot$ & $M_h/M_\odot$ & $\Omega\,M$ & $\bar m_p$ &$\bar \mathcal M_0$  &$\bar\mathcal M_A$ & $\bar\rho_c$ & $C$ \cr
 \mr
 Q2 & 2 & 1.456 & 2.7  & 0.0319 & 0.2733 & 0.1549 & 0.1436 & 0.1381 & 0.1529\cr
 Q3 & 3 & 1.456 & 4.05 & 0.0318 & 0.4123 & 0.1553 & 0.1439 & 0.1391 & 0.1536\cr
 Q5 & 5 & 1.457 & 6.75 & 0.0318 & 0.6907 & 0.1557 & 0.1443 & 0.1401 & 0.1543\cr
\br
\end{tabular}
\caption{Initial configuration parameters: $q=M_h/M_*$ binary mass ratio,  $M_0$ is the rest mass of the NS, $M_h$ irreducible mass of the \bh{}, $\Omega M$ orbital frequency, $\bar m_p =  m_p\,\kappa^{1/2}$ bare or puncture mass of the \bh{}, $\bar \mathcal M_0 = \mathcal M_0/\kappa^{1/2}$ rest mass of the \ns{} in isolation, $\bar \mathcal M_A = \mathcal M_A/\kappa^{1/2}$ ADM mass of the \ns{} in isolation, $\bar\rho_c = \rho_c\,\kappa$, central density of the isolated \ns{}, and $C = \mathcal{M}_A/R_*$ compactness of the isolated \ns{}. For all cases, $\kappa = 93.65\,M_\odot^2$, $\Gamma = 2$, coordinate separation $9\,M$, and $M_* = 1.35\,M_\odot$.}
\label{tab:ID}
\end{center}
\end{table}

We use the moving box mesh refinement approach as implemented by \texttt{Carpet}~\cite{2016ascl.soft11016S}. The starting point in setting the grid structure and number of refinement levels is the number of points needed to resolve the \bh{} and the \ns{.} For the results in this work, we ensure that at the finest level both, the \bh{} and the \ns{,} are completely enclosed by a mesh with at least 100 points across. This translates to a grid-spacing of $\sim 225$ meters for the \ns{.} From the finest level up, we add coarse refinements until we reach a resolution with grid-spacing $\sim M$, which is a suitable resolution for \gw{} extraction. For the mass ratios in this study, the end result is 9 levels of refinement from the \bh{} up to the coarsest and 8 for the \ns{.}

\begin{figure}
\centering
\includegraphics[width=0.8\textwidth]{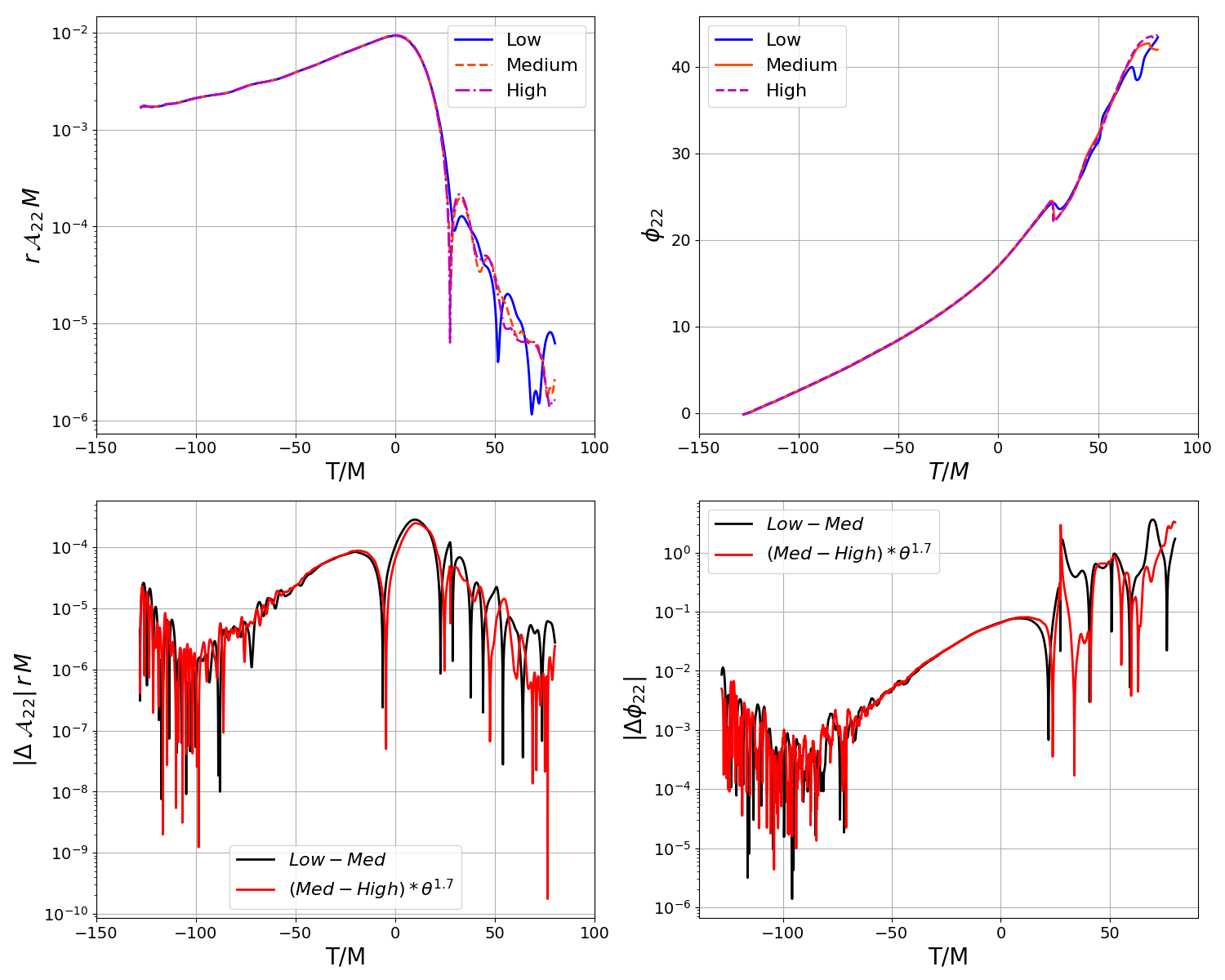}
\caption{Convergence results for the (2,2) mode of the Weyl scalar $\Psi_4$ for mass ratio $2$ case. Top panels show the amplitude (left) and phase evolution (right) for the three resolutions. Assuming a convergence rate of $k$ and refinement actor $\theta$, one should have that (Medium - Low) = $\theta^k$ (High - Medium). Bottom panels show the left and right hand side of this equation for amplitude (left) and phase (right) for $\theta = 1.5$ and $k = 1.7$.}
\label{fig:q2_psi4_convtest}
\end{figure}

\begin{figure}
\centering
\includegraphics[width=0.8\textwidth]{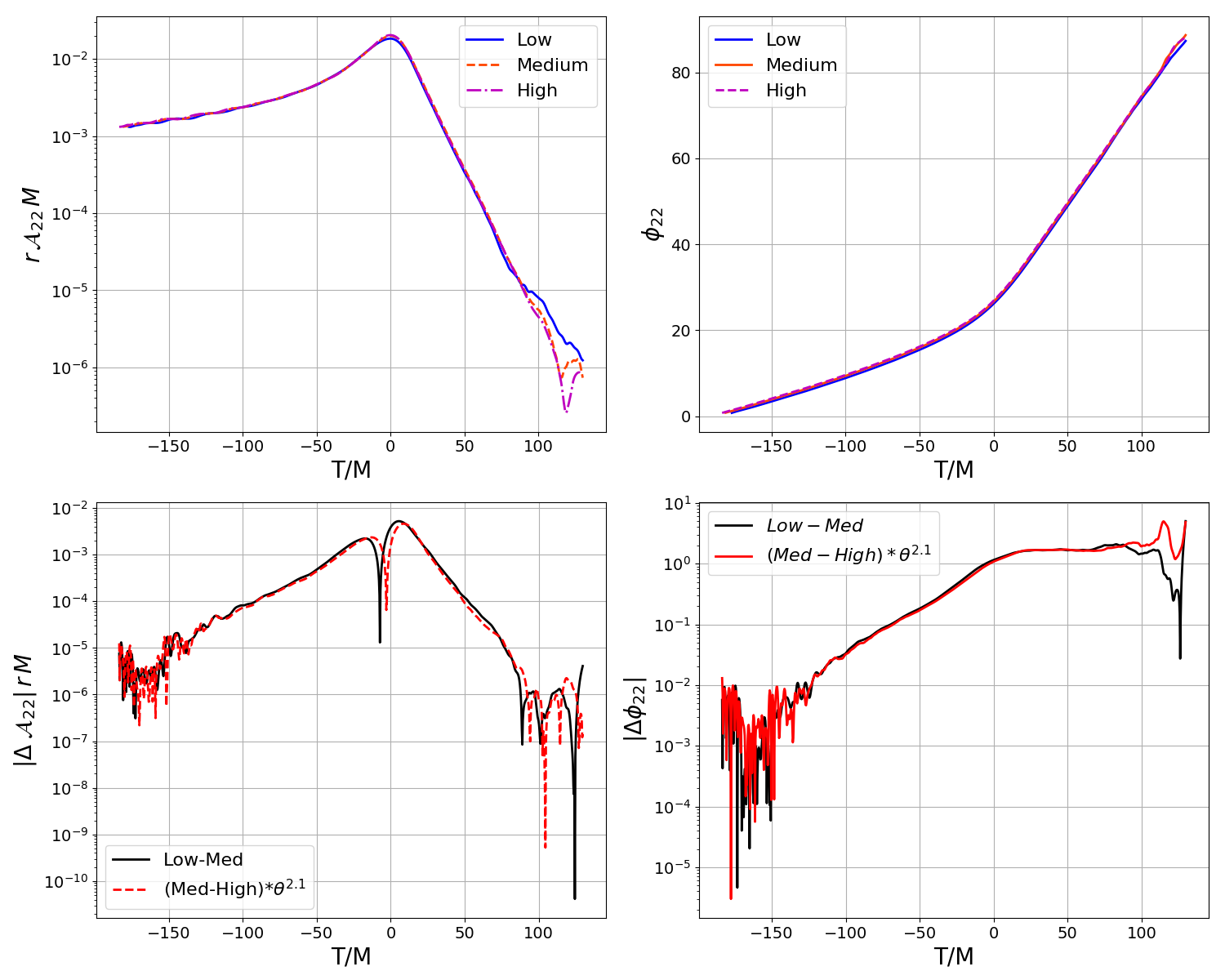}
\caption{Same as figure \ref{fig:q2_psi4_convtest} but for mass ratio $3$ case. Here, $\theta = 1.5$ and $k = 2.1$.}
\label{fig:q3_psi4_convtest}
\end{figure}

To test convergence, we carried out three simulations for $q=2$ at initial coordinate separation $7\,M$ with resolutions decreasing by a factor of $\theta = 1.5$. The finest meshes covering the \ns{} have grid-spacing $M/24$ (248 meters), $M/36$ (166 meters) and $M/54$ (102 meters). Since the size of the \bh{} is smaller than the \ns{,} the hole has an additional refinement level with twice the resolution of the finest mesh at the \ns{.} Figure~\ref{fig:q2_psi4_convtest} shows the convergence results for the (2,2) mode of the Weyl scalar $\Psi_4$. Top panels show the amplitude (left) and phase evolution (right) for the three cases. Assuming a  convergence rate of $k$ and refinement factor $\theta$, one should have that (Medium - Low) = $\theta^k$ (High - Medium). The panels on the bottom show the left and right hand side of this expression for $\theta = 1.5$ and $k = 1.7$. We see very good convergence in both amplitude and phase until $t \sim 25M$. Beyond this time, i.e. the late ringdown phase, convergence is not as clean. The high noise in the amplitude and phase differences in the bottom panels during the initial time is due to the junk radiation from the initial data. To understand the challenge of not having a clean convergence during the late rigndown, we carried out a convergence test for $q=3$ with resolutions of $M/21.43$ (374 meters), $M/32.14$ (248 meters) and $M/48.21$ (166 meters). Figure \ref{fig:q3_psi4_convtest} shows the convergence in amplitude and phase of (2,2) mode of $\Psi_4$ with $k=2.1$. We see now cleaner convergence even during the late ringdown phase. The main difference is that during the late ringdown for $q=3$ there is significantly less tidal debris being accreted by the \bh{.} Because of its low density, it is difficult to demonstrate clean convergence.

\begin{figure}[!htb]
\centering
\includegraphics[width=0.9\textwidth]{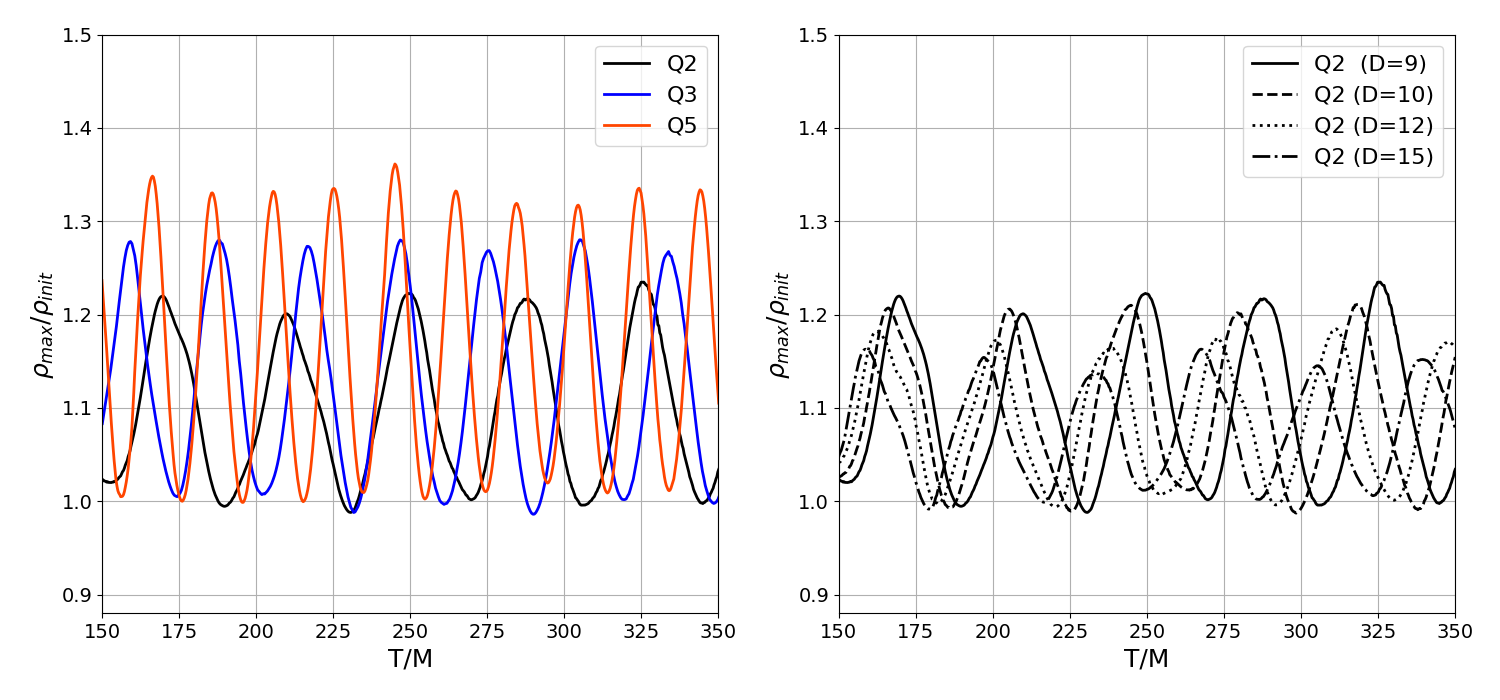}
        \caption[Density Oscillations]{Oscillations in the normalized central density of the \ns{}. Here, $\rho_{max}$ is the central density of the neutron star at any time $t$ and $\rho_{init}$ is the central density at initial time. Panel on the left shows a comparison of density oscillations for different mass ratio while panel on the right shows the comparison for four different separation for mass ratio 2.}
\label{fig:density_oscillations}
\end{figure}

A shortcoming of our initial data methodology is the presence of spurious oscillations in the \ns{.}  We suspect that the oscillations are likely triggered by how the star is boosted, i.e. via the Bowen-York extrinsic curvature, since we observe similar oscillations for tests with single boosted stars. Another source could be tidal effects, which are not included in the initial data. Although these effects could be reduced if one starts the simulations at larger separations.  In the left panel of Figure \ref{fig:density_oscillations}, we show the oscillations in the normalized rest mass density of the \ns{} as a function of time for the three mass ratio cases. We find oscillation amplitude ranges between 15 and 18\%, with the oscillation frequency that of the fundamental mode of the \ns{.} To test that the oscillations are likely due to the boost and tidal effects, we compare the oscillations for different initial separations for $q=2$, as shown in the panel on the right. As we increase the initial separation, tidal forces become weaker and the initial velocity of the star smaller, as a consequence the amplitude of the oscillations reduce.  We should stress that the oscillations are small and do not have any effect on the stability of the star.

\section{Results}
\label{sec:results}

\subsection{Inspiral and Merger Dynamics}

During the late inspiral stage of a \bhns{} binary,  the \ns{} will face a constant battle between the tidal forces from the \bh{} and its self gravity. Depending on the mass of the \bh{}, this could lead to the complete disruption of the star before it gets swallowed by the \bh{.} The tidal forces by the \ns{} could also inflict deformations in a companion, such as in a double \ns{} binary merger. However, although not generally accepted~\cite{2020arXiv201015795L}, there is strong evidence that \bh{s} are immune to tidal deformations~\cite{2020arXiv201007300C}. Thus, there are potentially fundamental differences between a \bbh{} and a \bhns{.}

\begin{figure}
\centering
\includegraphics[width=0.75\textwidth]{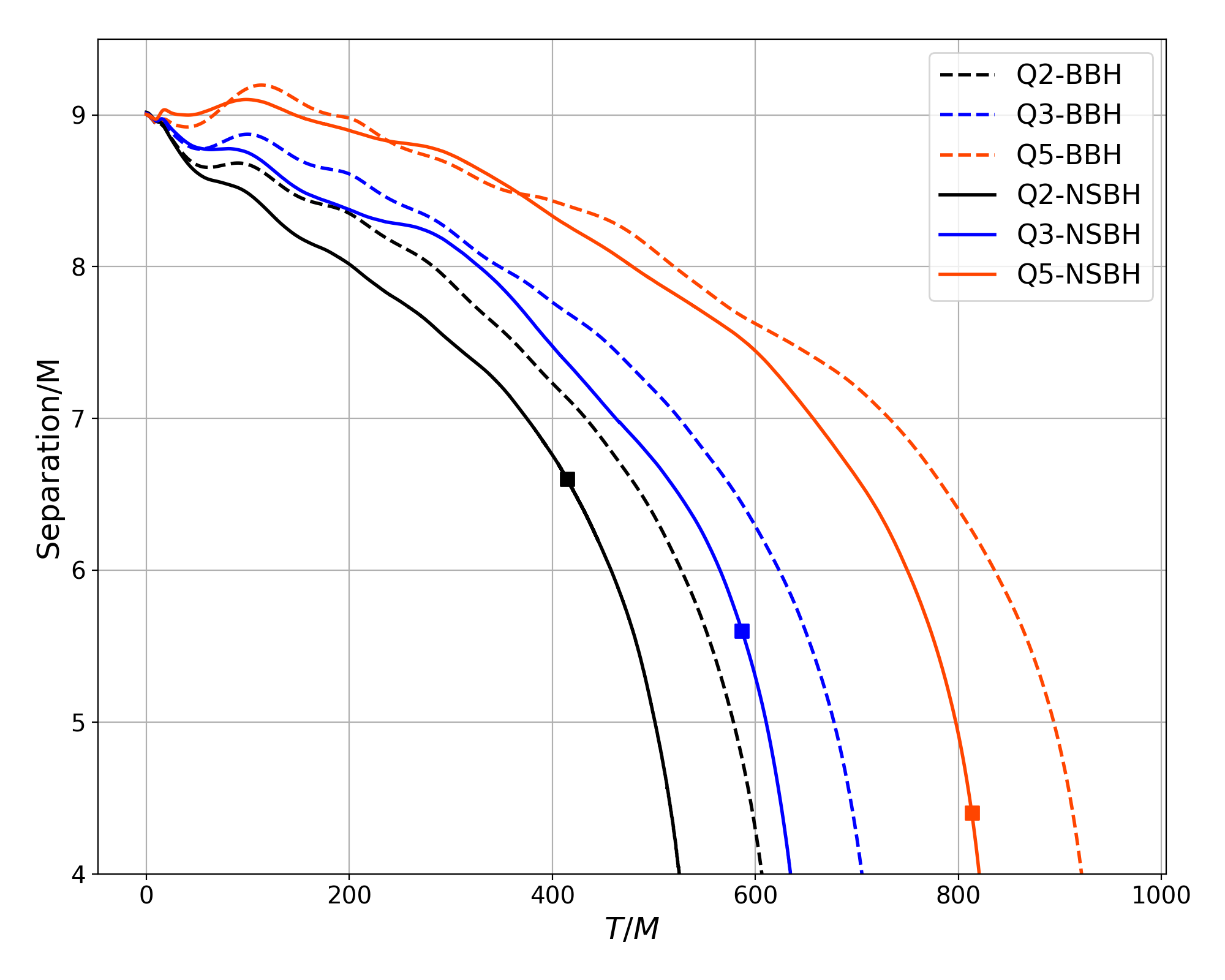}
\caption[Orbital Separation vs Time]{Evolution of the binary coordinate separation for each \bhns{} and \bbh{} binary described in Table \ref{tab:ID}. Solid squares denote when the binary has reached the tidal radius as computed by Eq.~\ref{eq:Rt}.}
\label{fig:sep}
\end{figure}

For compact object binaries, the luminosity of gravitational radiation and the rate of change of radiated angular momenta depend on the mass ratio as $q^2/(1+q)^4$~\cite{baumgarte_shapiro_2010}. As a consequence, the higher the mass ratio, the longer it takes for the binary to merge.
In Figure~\ref{fig:sep}, we show the evolution of the coordinate separation of the binary for each of the \bhns{} systems described in Table \ref{tab:ID} and its \bbh{} counterpart. The delay as a function of $q$ for both, the \bhns{} binaries and \bbh{s}, is evident in this figure. For a given $q$, we also see that the coordinate separation of the \bhns{} binary decreases faster than its corresponding \bbh{.} For the $q=5$ case, the \bbh{} and \bhns{} follow each other up to a separation $\sim 7.5\,M$. For $q=3$, the binaries diverge a little earlier at a separation of approximately $8\,M$. The earliest deviation occurs for the $q=2$ case, approximately at a time $50\,M$ from the start of the simulation. That is, as the mass ratio increases, the \bhns{} binaries resemble longer a \bbh{}. The difference in the binary separation between \bbh{} and \bhns{} systems grows stronger as the merger is approached.

Regarding the time when compact objects merge, for \bbh{}, it is marked by the sudden formation of a common apparent horizon. The apparent horizon appears a few $M$s before the gravitational radiation reaches peak luminosity. For \bhns{} binaries, we do not have the formation of a common apparent horizon since the only horizon is the one from the single \bh{} in the binary. Therefore, when making comparisons near coalescence, we will focus on the time when the gravitational radiation reaches peak luminosity (corresponding to the peak of $|\Psi_4|$).

In Table \ref{tab:table_orbdynamics}, $T_{mx}$ denotes the retarded time to peak luminosity and $\Delta_{mx}$ the final \bh{} offset  at $T_{mx}$.
 Notice that the \bhns{} binaries reach peak luminosity earlier than their corresponding \bbh{}.
 We will address the reasons for this difference when we discuss the \gw{s} emitted by the binaries. An interesting aspect to point out is that this difference does not decrease monotonically with  $q$. 

Also in Figure~\ref{fig:sep}, denoted with solid squares is the coordinate separation when the binary reaches the tidal radius as estimated by  \cite[Eq.17.19]{baumgarte_shapiro_2010}
\begin{equation}
    \frac{R_T}{M_h} \simeq 2.4\,q^{-2/3}\, C^{-1} \,.
\end{equation}
Relative to the ISCO radius $R_I \simeq 6\,M_h$, the tidal radius is given by
\begin{equation}
    \frac{R_T}{R_I} \simeq \left(\frac{q}{4.3}\right)^{-2/3}\left(\frac{C}{0.15}\right)^{-1}\,. \label{eq:ch2_iscoradius}
\end{equation}
For
$q\ge 4.3$, $R_T \le R_I$, and the \ns{} is swallowed by the \bh{} relatively intact.
For reference, the tidal radius in units of the total mass $M$ is given by
\begin{equation}
    \frac{R_T}{M} \simeq 16 \, \frac{q^{1/3}}{1+q}\left(\frac{C}{0.15}\right)^{-1}\,. \label{eq:Rt}
\end{equation}
Thus,
$R_T/M \simeq 6.6,\,5.6$ and $4.4$ for $q = 2,\,3$ and $5$, respectively. For reference,
the coordinate separation at the beginning of the simulations is $9\,M$

\begin{table}
\begin{center}
\begin{tabular}{@{}*{7}{l}}
\br
System & $q$ & $E_{ADM}/M$  & $J_{ADM}/M^2$ & $e/10^{-3}$ & $T_{mx}/M$ & $\Delta_{mx}/M$\cr
\mr %
\bbh{}  & 2 & 0.9901 & 0.829  &  6.3    & 648 & 0.027  \cr
\bhns{} & 2 & 0.9914 & 0.829  & 6.8     & 537 & 0.746   \cr
\bbh{}  & 3 & 0.9918 & 0.702  & 5.5     & 743 & 0.058   \cr
\bhns{} & 3 & 0.9921 & 0.702  & 8.3     & 662 & 0.782   \cr
\bbh{}  & 5 & 0.9939 & 0.523  & 10.5    & 957 & 0.037   \cr
\bhns{} & 5 & 0.9941 & 0.523  & 9       & 848 & 1.03   \cr
\br
\end{tabular}
\caption{ Binary system, mass ratio $q$, ADM energy $E_{ADM}$ and angular momentum $J_{ADM}$, eccentricity $e$, retarded time to peak luminosity $T_{mx}$, and final \bh{} offset $\Delta_{mx}$ at $T_{mx}$.}
\label{tab:table_orbdynamics}
\end{center}
\end{table}

\begin{figure}[!htb]
\centering
\includegraphics[width=\textwidth]{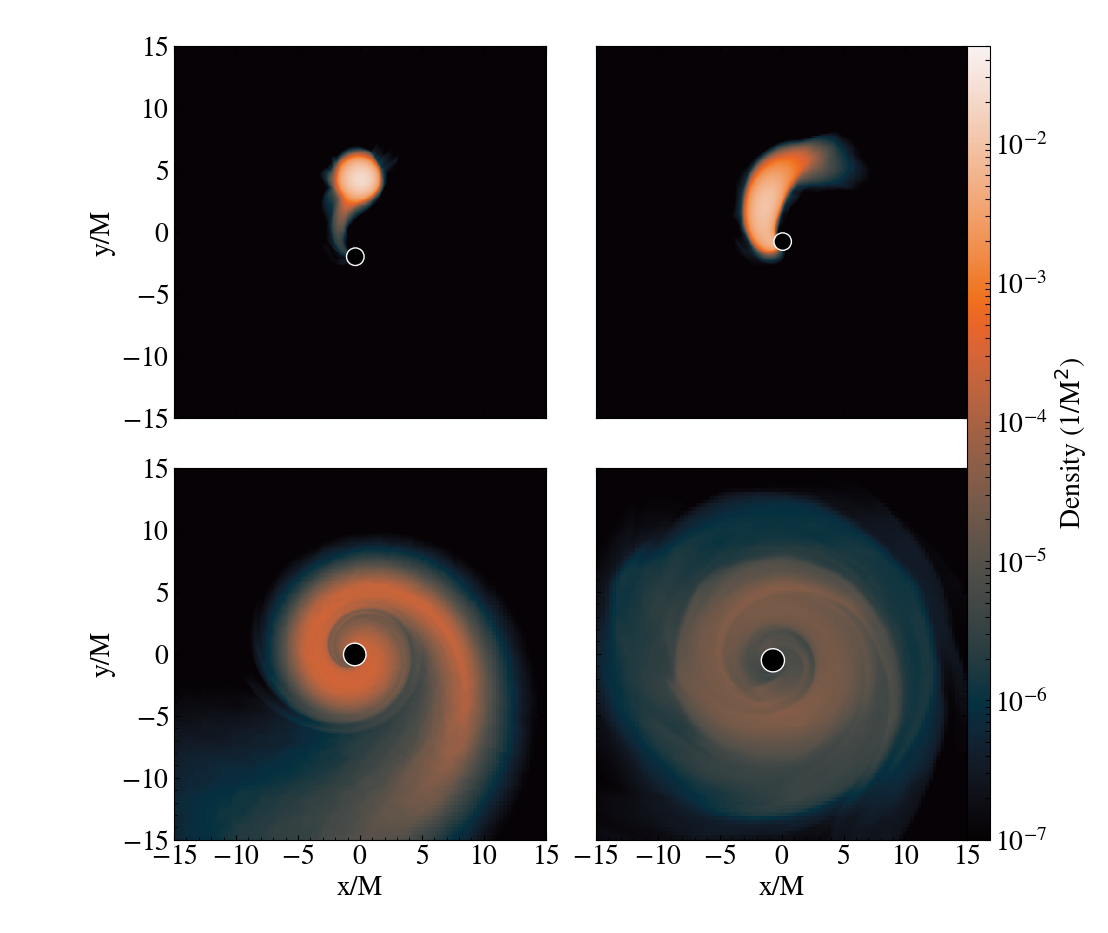}

        \caption[Rest Mass Density contour plots for $q=2$]{Snapshots of the rest mass density for the $q=2$ \bhns{.} The \bh{} is represented by a black circle with white boundary. Panel on the top left shows the beginning of stellar disruption $96\,M$ before peak luminosity. Top right panel shows the stellar disruption at the time of merger followed by circularization of matter forming an spiral arm around \bh{} $100\,$M after the merger (bottom left panel). The last panel shows the final state of the accretion disk 500\,M after the merger.}
\label{fig:q2_densityplots}
\end{figure}

\begin{figure}[!htb]
\centering
\includegraphics[width=\textwidth]{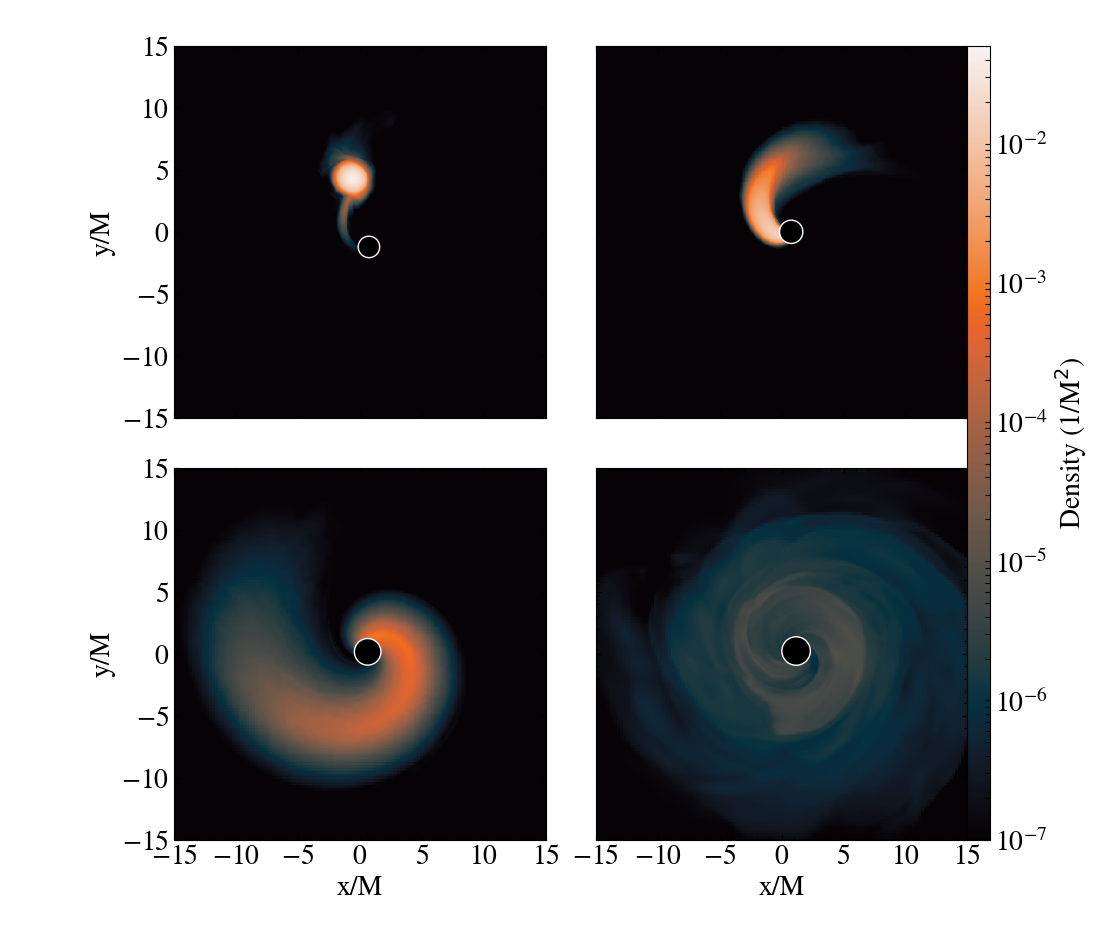}

        \caption[Rest Mass Density contour plots for the $q=3$]{Snapshots for  $q=3$ \bhns{} merger. Top left panel shows the beginning of tidal disruption and tail formation $78\,M$ before peak luminosity, and the top right panel shows the consumption of disrupted star by the \bh{} at peak luminosity. The bottom left panel shows matter circularization $30\,M$ after peak luminosity. The bottom right panel depicts the formation of a very tenuous accretion disk $500\,M$ after peak emission.  }
\label{fig:q3_densityplots}
\end{figure}

\begin{figure}[!htb]
\centering
\includegraphics[width=\textwidth]{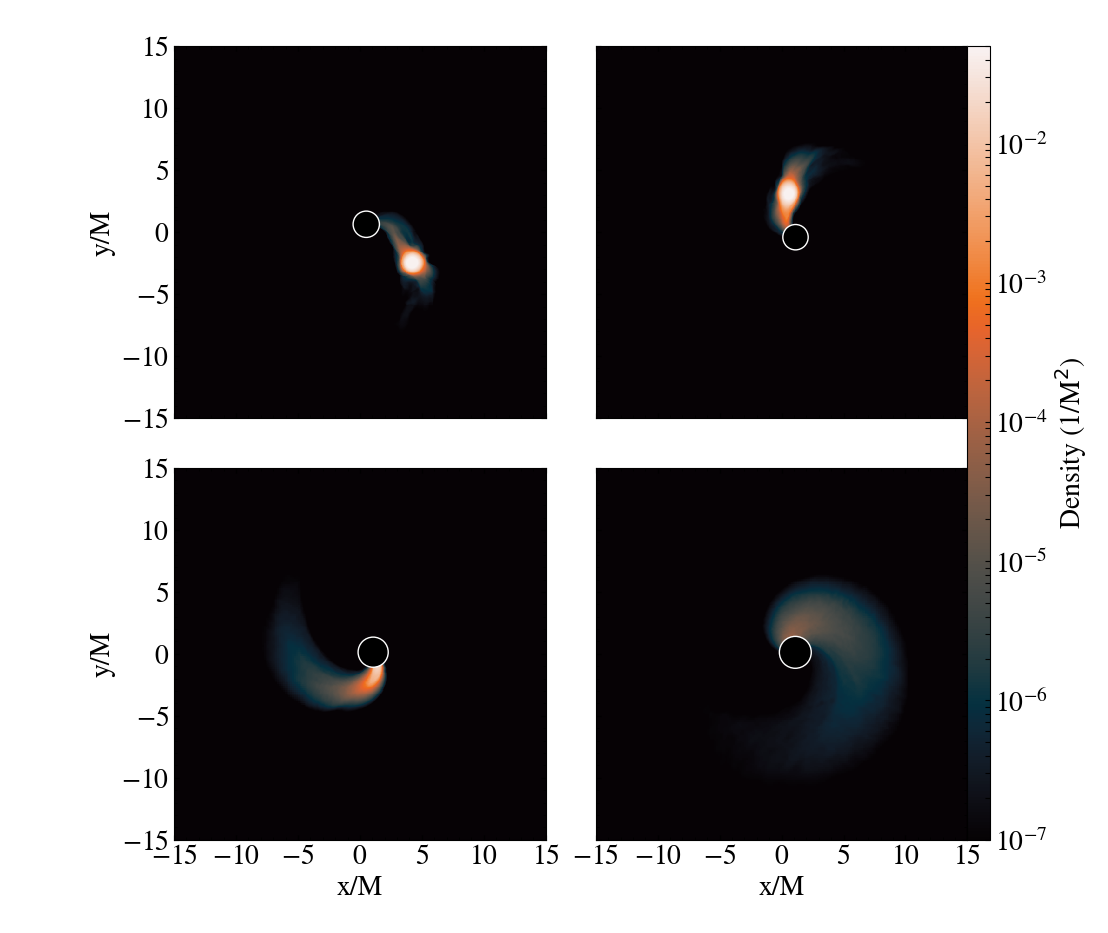}

        \caption[Rest Mass Density contour plots for $q=5$]{Snapshots for the \bhns{} $q=5$ binary merger. The top left panel shows the rest mass density at $45\,M$ prior peak luminosity. The top right panel shows the situation $20\,M$ before peak luminosity and the bottom left panel at the merger. The bottom right panel shows the result $20\,M$ after peak luminosity.}
\label{fig:q5_densityplots}
\end{figure}

To get an overall sense of the inspiral and merger, Figures~\ref{fig:q2_densityplots}, \ref{fig:q3_densityplots} and \ref{fig:q5_densityplots} show snapshots of the rest mass density in the orbital plane for all the cases under consideration.
For $q=2$, the tidal forces from the \bh{} trigger mass shedding early on, at approximately $440\,M$ from the beginning of the simulation when the binary separation is approximately $6.25\,M$. This happens roughly $96\,M$ before the peak luminosity.
Figure~\ref{fig:q2_densityplots} shows four evolution snapshots for this case. The \bh{} is represented by a black circle with white boundary. The initial central density of star is $0.02 \,M^{-2}$ ($7.7 \times 10^{14}~\textrm{g/cm}^3$). Top left panel shows a snapshot at time  $96\,M$ before the peak luminosity, when the \ns{} begins to be disrupted. Top right panel shows the stellar disruption at the time of merger. Notice that the \ns{} has been completely destroyed, deforming into a spiral arm around the \bh{} which extends to $7\,M$ beyond the hole.  Bottom left panel show the circularization stage of matter around the \bh{} about 100\,M after the merger. We found that about $90\%$ of the star's material falls into the \bh{} within the first $100\,M$ of evolution while the remaining material continues to expand outwards slowly morphing into an accretion disk. The bottom right panel shows the final state of the accretion disk 500\,M after the merger, reaching a core density of  $\sim 10^{-4}\,M^{-2} $ ($\sim 10^{12}\, \textrm{g/cm}^3$). 

The  $q=3$ \bhns{} merger follows the $q=2$ steps but not as dramatic in terms of disruption effects. In Figure \ref{fig:q3_densityplots}, the top left panel shows the beginning of tidal disruption and tail formation $78\,M$ before the merger. The channel of mass transfer is much narrower due to weaker tidal interactions. This is followed by complete disruption of the star at the merger shown in the top right panel more than $95\%$ of which is consumed by the \bh{} within $30\,M$ . The bottom left panel shows matter circularization $30\,M$ after the merger. The bottom right panel depicts the formation of a very tenuous accretion disk $500\,M$ after peak emission with characteristic density of $\sim 10^{-5}\,M^{-2}$ ($\sim 10^{11}\, \textrm{g/cm}^3$).

As mentioned before, the $q=5$ for a \bhns{} behaves more like a \bbh{,} with the star remaining almost intact by the time it reaches $R_I$ since $R_T \simeq 4.4\,M$. The top left panel in Figure \ref{fig:q5_densityplots} shows a snapshot at $45\,M$ prior to the merger. There are hints of material being stripped from outer layers of the star. The top right panel shows the situation $20\,M$ before the merger and the bottom left panel at the merger. The bottom right panel shows the result $20\,M$ after the merger. At that point, $99\%$ of the \ns{} has been swallowed by the hole.  This leaves a remnant state with extremely low densities. Since there is very little change in this case of triggering electromagnetic signatures, the \bhns{} and \bbh{} are almost indistinguishable from each other. This will be more apparent when we compare \gw{} emissions.


\begin{table}
\begin{center}
\begin{tabular}{@{}*{10}{l}}
 \br
System & $q$  & $M_h/M$ & $M_*/M$ & $M_f/M$ & $M_c/M$ & $M_{td}/M$ & $a_f$ & $E_{rad}/M$ & $v_k (\textrm{km/s})$\cr 
 \mr
\bbh{}  & 2  & 0.667 & 0.333 & 0.9073 & 0.9598  &          & 0.617 & 0.0295  & 147.7  \cr
\bhns{} & 2  & 0.667 & 0.333 & 0.8984 & 0.9659  & 0.019   & 0.683 & 0.0077 & 39.1 \cr
\bbh{}  & 3  & 0.750 & 0.250 & 0.9319 & 0.9712  &          & 0.5405 & 0.0208  & 169 \cr
\bhns{} & 3  & 0.750 & 0.250 & 0.9308 & 0.974   & 0.008   & 0.563 & 0.0102 & 29.2  \cr
\bbh{}  & 5  & 0.833 & 0.167 & 0.9598 & 0.9823  &          & 0.4166 & 0.0118 & 135.7\cr
\bhns{} & 5  & 0.833 & 0.167 & 0.9603 & 0.9834  & 0.0004   & 0.4203 & 0.0106 & 103.5 \cr
 \br
\end{tabular}
        \caption{$M_h$ irreducible mass of the initial \bh{} (mass of the larger \bh{} in \bbh{} cases), $M_*$ initial mass of the \ns{} (irreducible mass of the smaller \bh{} in \bbh{} cases), $M_f$ irreducible mass of the final \bh{}, $M_c$ Christodoulou mass of the final \bh{}, $M_{td}$ mass left outside the final \bh{}, $a_f$ dimensionless spin of the final \bh{}, $E_{rad}$ radiated energy, and $v_k$ its kick.}\label{tab:table_remnant}
\end{center}
\end{table}

\subsection{The Final \bh{}}

The mass and spin of the final \bh{} in a \bhns{} merger will depend on the extent to which the \ns{} is devoured by the \bh{}. Table \ref{tab:table_remnant} shows $M_h$ the irreducible mass of the initial \bh{} (mass of the larger \bh{} in \bbh{} cases), $M_*$ the initial mass of the \ns{} (irreducible mass of the smaller \bh{} in \bbh{} cases), $M_f$ the irreducible mass of the final \bh{}, $M_c$ the Christodoulou mass of the final \bh{}, $M_{td}$ the mass left outside the final \bh{}, $a_f$ the dimensionless spin of the final \bh{}, $E_{rad}$ the radiated energy, and $v_k$ the kick of the final \bh{}.

First thing to notice is that the irreducible and Christodoulou masses of the final \bh{} in the \bhns{} and \bbh{} are comparable. On the other hand, the energy radiated and the spins and kicks of the final \bh{} differ significantly. \bhns{} mergers produce a final hole with higher spin but with a lower kick. The differences in both the final spin and kick decrease as $q$ increases since the binary becomes more \bbh{-}like. The main culprits of the differences are again the tidal deformations and disruption of the \ns{.}

\begin{figure}
\centering
\includegraphics[width=0.85\textwidth]{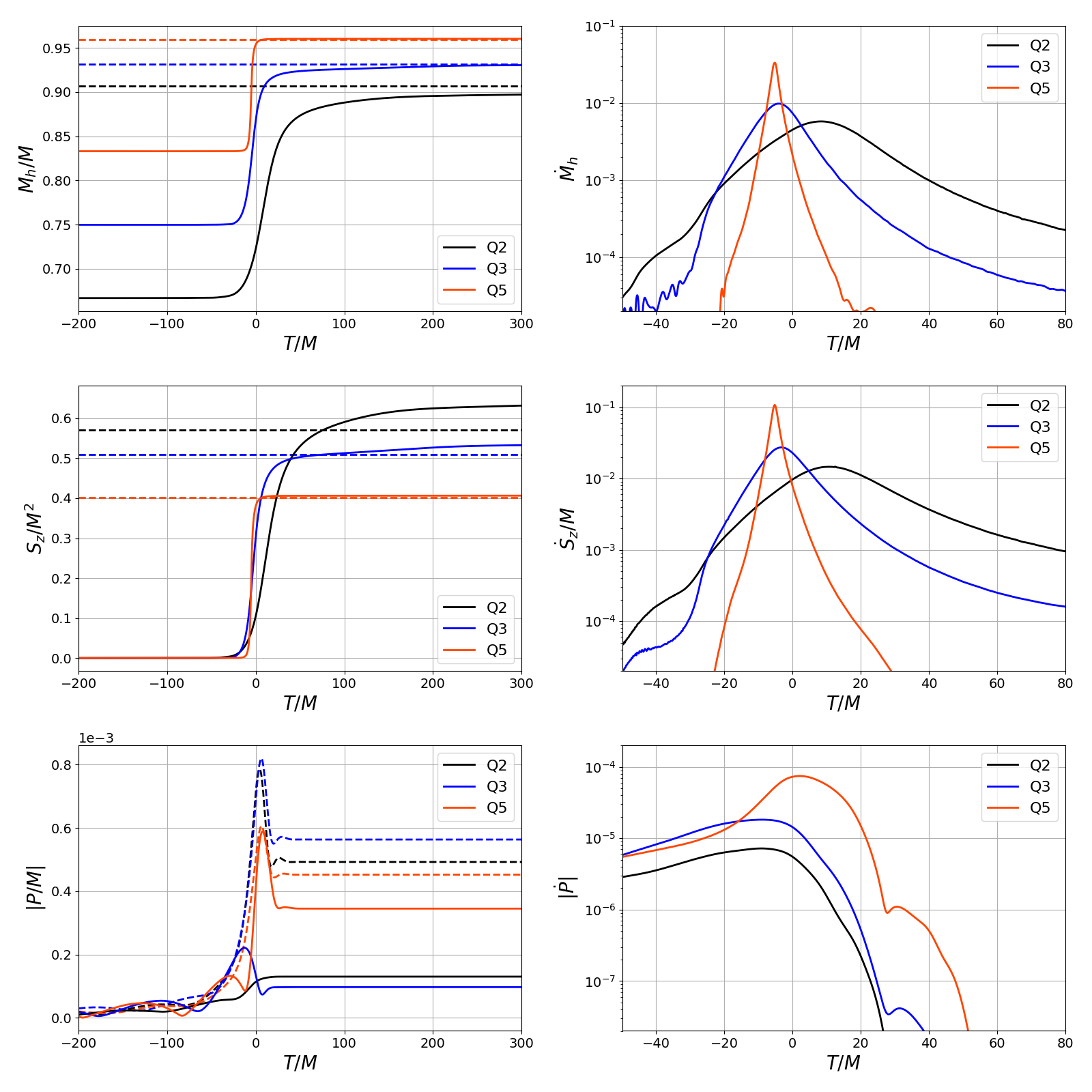}
        \caption{Top left panel shows with solid lines the growth of the irreducible mass of the \bh{} as a function of time for \bhns{} binaries. Dashed horizontal lines denote the final mass of \bh{} for the corresponding \bbh{} merger. $T=0$ is the time at peak luminosity. Top right panel shows the corresponding mass growth rate $\dot M_h$. Middle left panel shows with solid lines the growth of the spin of the \bh{} for \bhns{} binaries. For reference, dashed horizontal lines denote the spin of the final \bh{} for the corresponding \bbh{} merger. Middle right panel shows the corresponding spin growth rate $\dot S_z$. Bottom left panel shows the accumulation of linear momentum emitted by \gw{s} for both, the \bhns{} (solid lines) and \bbh{} systems (dashed lines). Bottom right panel depicts the corresponding rate of linear momentum accumulation.}

\label{fig:BH_mass}
\end{figure}

To understand the differences in the mass, spin and kicks of the final \bh{}, we plot in Figure \ref{fig:BH_mass} their evolution. The top left panel shows with solid lines the growth of the irreducible mass of the \bh{} in \bhns{} mergers. Dashed horizontal lines denote the final mass of \bh{}, $M_f$, for the corresponding \bbh{} merger. $T=0$ is the time at the peak luminosity. Notice that as expected, for $q=5$, the growth is abrupt because the \ns{} is swallowed almost intact, thus mimicking a \bbh{} in which a common apparent horizon suddenly appears to signal the merger. For $q=2$ the transitions takes much longer and the final mass of the \bh{} does not get closer to the mass of the \bbh{} final \bh{}. This is because of the material left behind. The rates at which the mass of the final \bh{} changes are depicted in the top right panel of Figure~\ref{fig:BH_mass}. The rates clearly emphasize that the growth is sharp for $q=5$ and smoother for $q=2$.

Regarding the spin of the final \bh{,} the middle left panel in Figure~\ref{fig:BH_mass} shows with solid lines the growth of the spin of the final \bh{} for \bhns{} binaries given by $S_z/M^2$, and, for reference, dashed horizontal lines denote the spin of the final \bh{} for the corresponding \bbh{} merger. Middle right panel shows the corresponding spin growth rate $\dot S_z/M$. Here also one observes that for lower $q$ the transition is smoother. Important to notice that $S_z/M^2$ is not the dimensionless spin of the final \bh{}. The dimensionless spin is given $a_f = S_z/M_c^2$ with $M_c$ the Christodoulou mass of the final \bh{}. The reason why $a_f$ for \bhns{} are higher is because, as we will see later, the emission of gravitational radiation carrying out angular momentum is lower; thus, at merger, the final \bh{} is left with high angular momentum.

For the kicks of the final \bh{} the situation reverses. The gravitational recoil is lower for \bhns{} mergers. This is because most of the accumulation of the gravitational recoil in compact object binaries takes place in the last few orbits, but this is precisely the stage when \bbh{} and \bhns{} differ the most. As the \ns{} undergoes disruption, and thus lose its compactness, the \bhns{} binary radiates less and with it the opportunity to carry out linear momentum. This is clear from the bottom panels in Fig.~\ref{fig:BH_mass} where the left panel shows the accumulation of linear momentum emitted by \gw{s} for both, the \bhns{} (solid lines) and \bbh{} systems (dashed lines).
It is interesting to notice that while the magnitude of the kicks for \bbh{s} are $Q5<Q2<Q3$, consistent with the results in Ref.~\cite{2007PhRvL..98i1101G}, the kicks for \bhns{} systems are $Q3<Q2<Q5$.

\begin{figure}[!htb]
\centering
\includegraphics[width=0.7\textwidth]{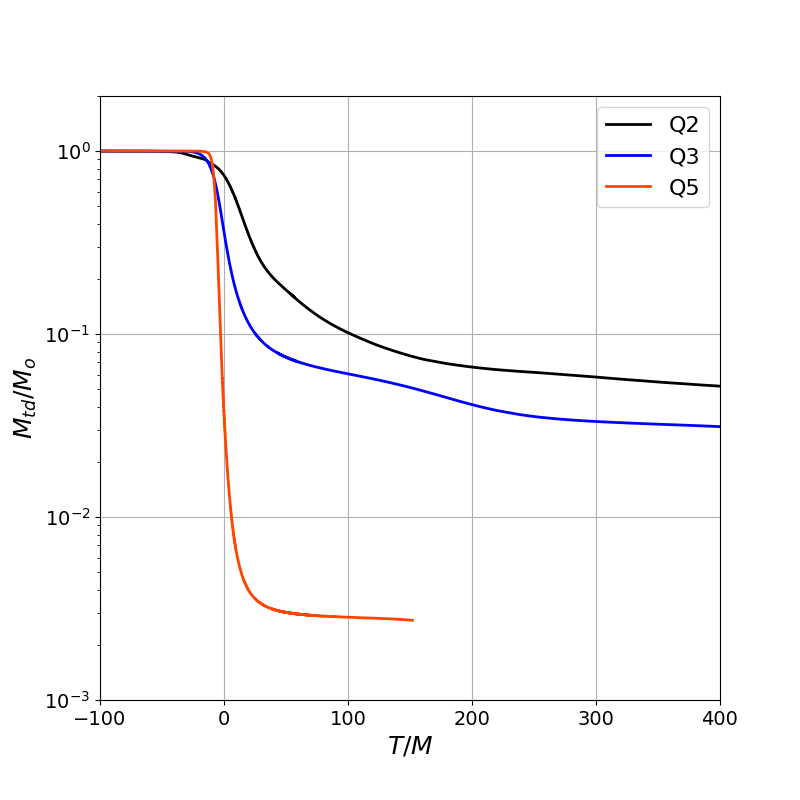}

        \caption[Rest Mass Outside BH]{Rest mass of the tidal debris $M_{td}$ outside the horizon normalized to the initial rest mass of the \ns{}.}
\label{fig:restmass_outside_bh}
\end{figure}

\subsection{Tidal Debris}

We have seen from Figure~\ref{fig:q2_densityplots} that the disruption of the \ns{} would leave behind a trail of material in the vicinity of the \bh{}.  To get a better understanding of how the remnant material outside the \bh{} depends on the mass ratio $q$, we plot in Figure~\ref{fig:restmass_outside_bh} the rest mass of tidal debris $M_{td}$ outside the \bh{} normalized to the initial rest mass of the \ns{.} After peak luminosity, $M_{td}$ accounts for both the mass in the accretion disk and the unbound debris. For $t < -50\,M$, the material from the \ns{} has not reached yet the hole; thus, $M_{td}$ includes the entire mass of the star. 

For $q=2$, approximately $90\%$ of the mass of the \ns{} falls into the \bh{} within $t \sim 100\,M$ after peak luminosity. By $t \sim 400\,M$, the accretion process slows down, leaving behind $~5$\% of the mass of the \ns{}. The $q=3$ case follows a similar trend. At $t \sim 100\,M$, the \bh{} has already consumed $\sim 94\%$ of the stellar material, and at $t \sim 400\,M$ only $\sim 3\%$ is left outside the holes. As expected, the case $q=5$ is significantly different since $99\%$ of star is devoured by the hole just  $t\sim 20\,M$ after peak luminosity, leaving outside barely any material. Table \ref{tab:table_remnant} lists the final $M_{td}$ for each case. Comparing with the results in \cite{Shibata:2009cn} and \cite{Etienne:2008re}, we find that our masses agree well for $q=2$ but they differ by approximately $20\%$ for $q=3$. A possible explanation for this difference could be the effects from the artificial atmosphere used in this type of simulations to handle the vacuum regions in the computational domain. 

\subsection{Gravitational Waves}

\begin{figure}
\centering
\includegraphics[width=0.85\textwidth]{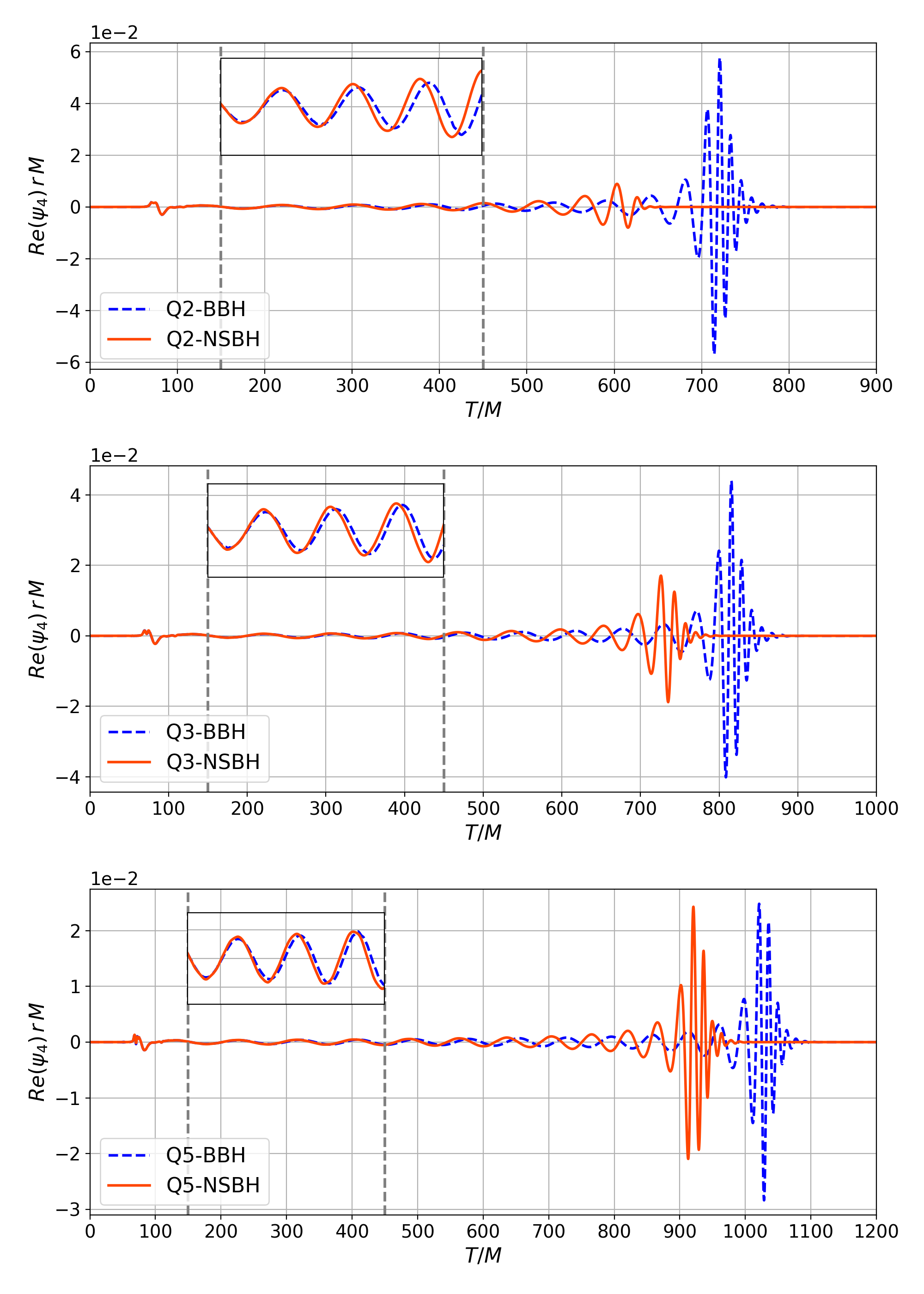}
        \caption{Real part of the (2,2) mode of the Weyl scalar $\Psi_4$. From top to bottom $q=2,\,3$ and 5. \bhns{} waveforms are depicted with solid lines and the corresponding \bbh{} waveform with dashed lines. The inset shows the waveforms early on between $150 \le T/M \le 450$.}
\label{fig:psi4}
\end{figure}

\begin{figure}
\centering
\includegraphics[width=0.85\textwidth]{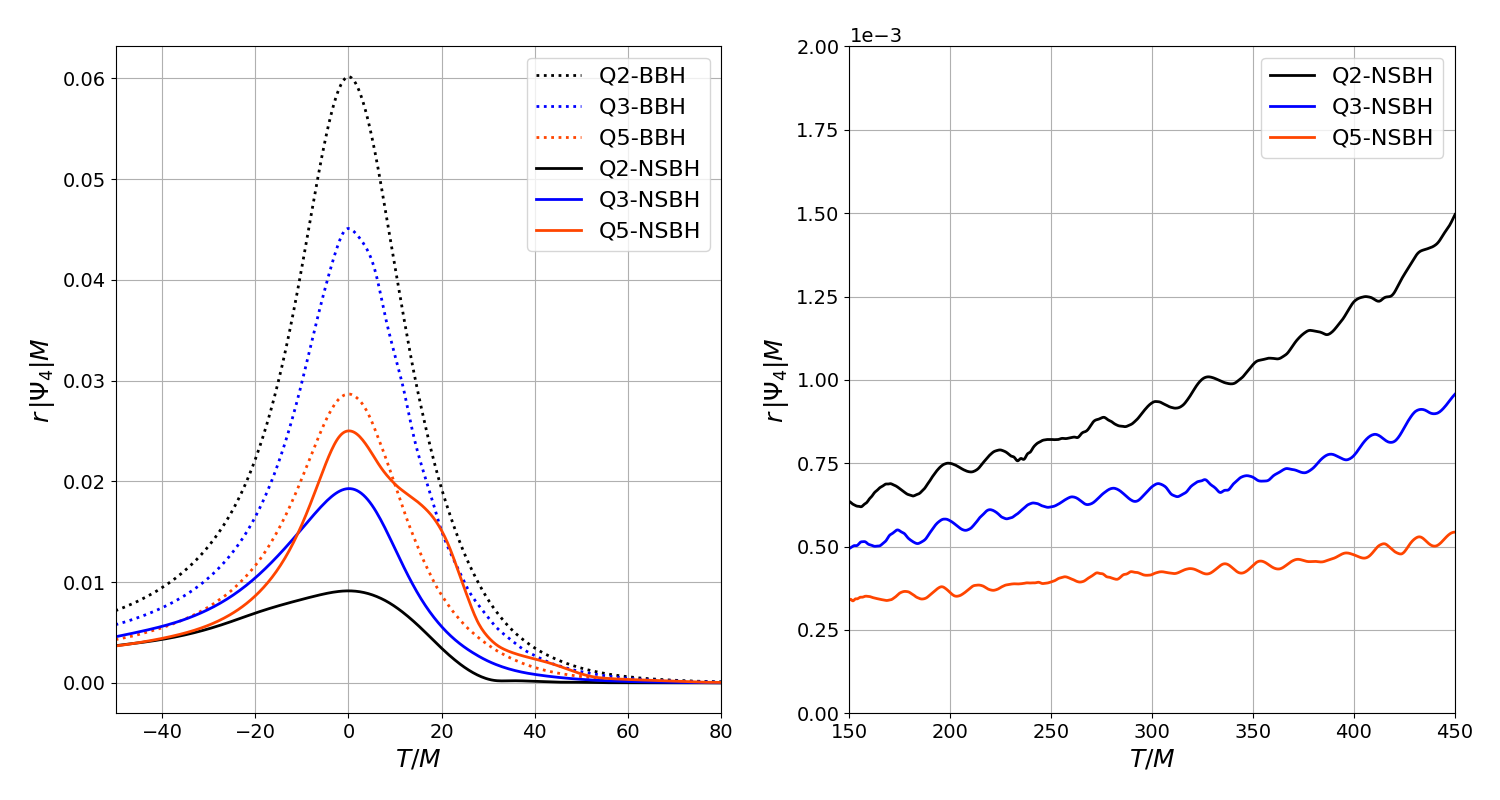}

        \caption[$\Psi_4$ Plots]{Amplitude of the (2,2) mode of $\Psi_4$. The left panel shows the waveform amplitudes of both the \bbh{} and \bhns{} mergers around peak luminosity. The right panel shows the waveform amplitudes for \bhns{} binaries during the time window of the insets in Figures~\ref{fig:psi4}.}
\label{fig:psi4amp}
\end{figure}

\begin{figure}
\centering
\includegraphics[width=0.85\textwidth]{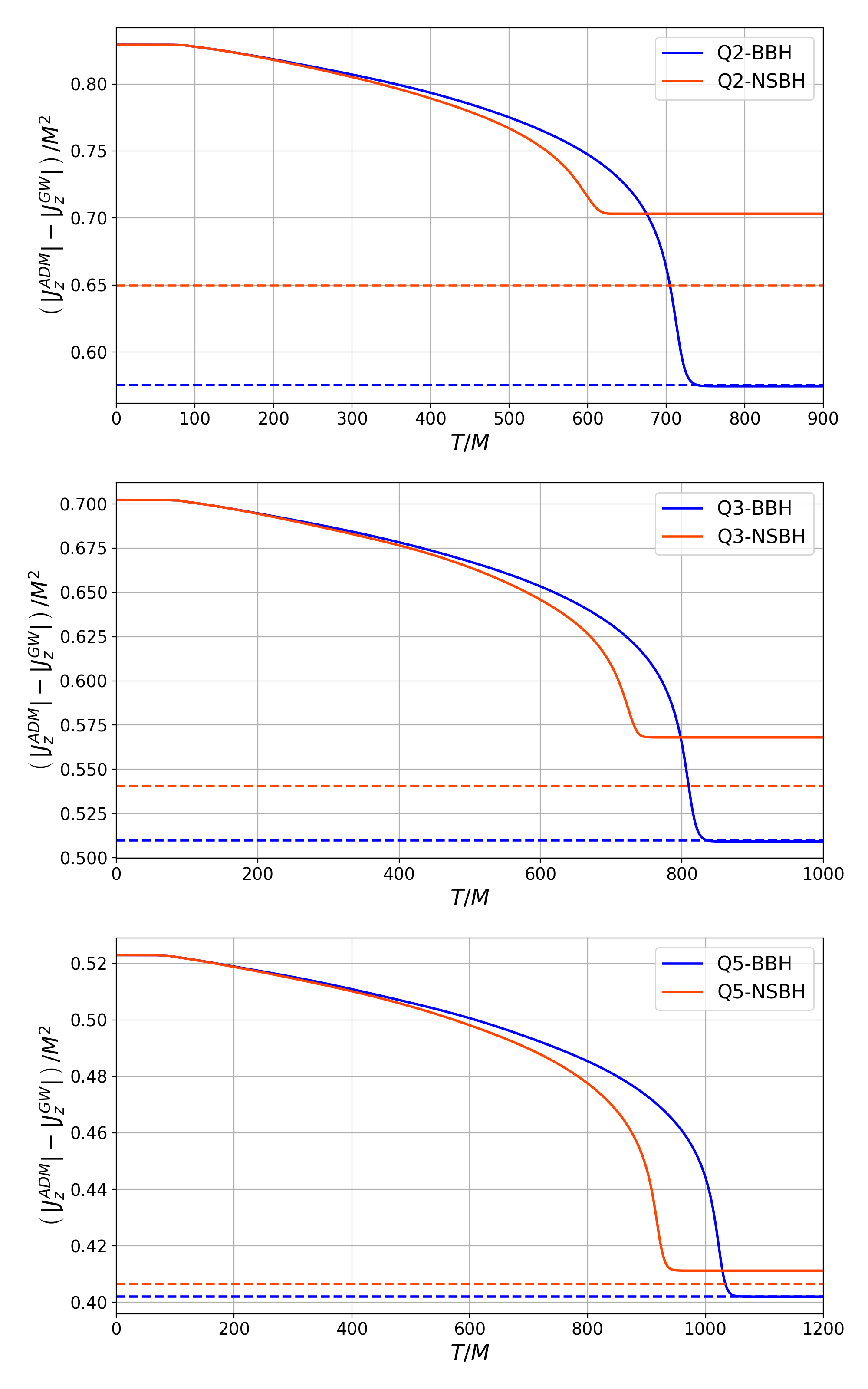}
        \caption{$|J_z^{ADM}|-|J_z^{GW}|$ with $J_z^{ADM}$ is the ADM angular momentum in the initial data and $J_z^{GW}$ the angular momentum carried out by \gw{s}. Dashed lines denote the angular momentum of the final \bh{}. For both, dashed and solid lines, blue denotes the \bbh{} case and red the \bhns{}. From top to bottom the cases $q=2$, 3 and 5, respectively.}
\label{fig:Jz}
\end{figure}

Figure \ref{fig:psi4} shows the real part of the (2,2) mode of the Weyl scalar $\Psi_4$ for the \bhns{} binaries (solid line) together with their corresponding waveform for the \bbh{} (dashed line). The insets show the waveforms early on between $150 \le T/M \le 450$, from which one can see that the \bbh{} and \bhns{} waveforms are closer to each other as $q$ grows.
In this figure, it is also evident that $\Psi_4$ for \bhns{} binaries reaches its maximum amplitude earlier. Since peak luminosity also signals that the binary merges around that time, this also implies that \bhns{} binaries merge earlier than their corresponding \bbh{} system. Before addressing the reasons for the prompt merger of \bhns{} binaries, we will discuss the differences in the peak luminosity.

Figure \ref{fig:psi4amp} depicts the amplitude of $\Psi_4$. The left panel shows the waveform amplitudes of both the \bbh{} and \bhns{} mergers around peak luminosity. For the \bbh{} amplitudes, from highest to lowest are $q=2,\,3$ and 5, respectively. This is not surprising since, as stated before, the luminosity of gravitational radiation depends on the mass ratio as $q^2/(1+q)^4$. Interestingly, the situation reverses for \bhns{} systems. The amplitudes are not only lower than those of the \bbh{s}, but now instead from highest to lowest are $q=5,\,3$ and 2. For $q=2$ and 3, the decrease in amplitude is due to the disruption the \ns{} experiences that makes it loose compactness and thus decrease the quadrupole moment of the binary. The $q=5$ \bhns{} case is comparable to the \bbh{} case because, once again, this is the case in which the star merges with the hole without significant disruption. The ``bump'' observed in this case is an artifact of the way the spherical decomposition is done. It assumes that the coordinate system is centered at the origin of the computational domain. From Table \ref{tab:table_orbdynamics}, we see that at merger time the center of mass of the binary for $q=5$ is already displaced $1\,M$ from the origin. As a consequence the (2,2) mode has contributions from higher modes. The other two $q$ cases also undergo displacements, but they are not as large, and the higher modes for low $q$'s are not as dominant as in $q=5$.

The right panel in Figure \ref{fig:psi4amp} shows the amplitudes of $\Psi_4$ for \bhns{} binaries during the time window of the insets in Figure \ref{fig:psi4}. Since this is during the early stage of the simulation, disruption effects do not play a significant role yet, and the situation resembles the \bbh{} case in which the amplitude decreases with $q$. The oscillations in the amplitude of $\Psi_4$ are due to the spurious oscillations in the \ns{} because of the initial data. In separating amplitude and phase in $\Psi_4$, we did not explicitly separated the effects from the \ns{} oscillations, thus, the oscillatory behavior remains in the amplitude.

Regarding the prompt merger of \bhns{} systems, there are only two channels to transport angular momentum out of the binary to harden it. As with \bbh{} systems, one channel is  via \gw{} emission. The other channel is transport of angular momentum by the tidal debris. Because the initial data is constructed using a generalization of the puncture Bowen-York approach, the ADM angular momentum in the initial data is the same for the \bhns{} binary and its corresponding \bbh{} as stated in Table~\ref{tab:table_orbdynamics}. Since we are dealing with non-spinning \bh{s} and \ns{s}, we only need to look at the angular momentum perpendicular to the orbital plane, namely the $z$-component. In Figure~\ref{fig:Jz}, we plot with solid lines $|J_z^{ADM}|-|J_z^{GW}|$ from top to bottom the cases $q=2$, 3 and 5, respectively. Here, $J_z^{ADM}$ is the ADM angular momentum in the initial data, and $J_z^{GW}$ is the angular momentum carried out by \gw{s}. In Figure~\ref{fig:Jz}, dashed lines denote the angular momentum of the final \bh{}. For both, dashed and solid lines, blue denotes the \bbh{} case and red the \bhns{}. Since for \bbh{s} there is only one channel, \gw{s}, to remove angular momentum from the binary, $|J_z^{ADM}|-|J_z^{GW}|$ after merger and ring-down closely matches the value of the angular momentum of the final \bh{.} The slight difference is because of the junk radiation in the initial data.

The situation is different for \bhns{} binaries. The first thing to notice in Figure~\ref{fig:Jz} is that there is a gap between the value that $|J_z^{ADM}|-|J_z^{GW}|$ reaches after merger and ring-down and the value of the angular momentum of the final \bh{.} This gap is closed if in addition one includes the angular momentum carried out by the tidal debris. The gap is larger the lower the $q$ because the tidal disruptions is stronger. The other feature in Figure~\ref{fig:Jz} is that the decrease of $|J_z^{ADM}|-|J_z^{GW}|$ is faster for \bhns{} binaries. The differences start appearing after approximately $300\,M$, $400\,M$ and $500\,M$ of evolution for $q=2$, 3 and 5, respectively. At those times, as clear from equation~\ref{eq:Rt}, the binary is far from the tidal disruption separation.
Therefore, the most likely culprit is the tidal deformations on the \ns{}. This effect was pointed out in Ref.~\cite{Shibata:2011jka} using post-Newtonian arguments. Specifically, it was noted that
the deformation in the \ns{} introduces a correction term in the potential whose magnitude increases
steeply with the decrease of the orbital separation. The effect is an acceleration of the inspiral and thus on the emission of \gw{s}, leading to a prompt merger.

\subsection{Quasi-normal Ringing}

\begin{figure}
\centering
\includegraphics[width=1.0\textwidth]{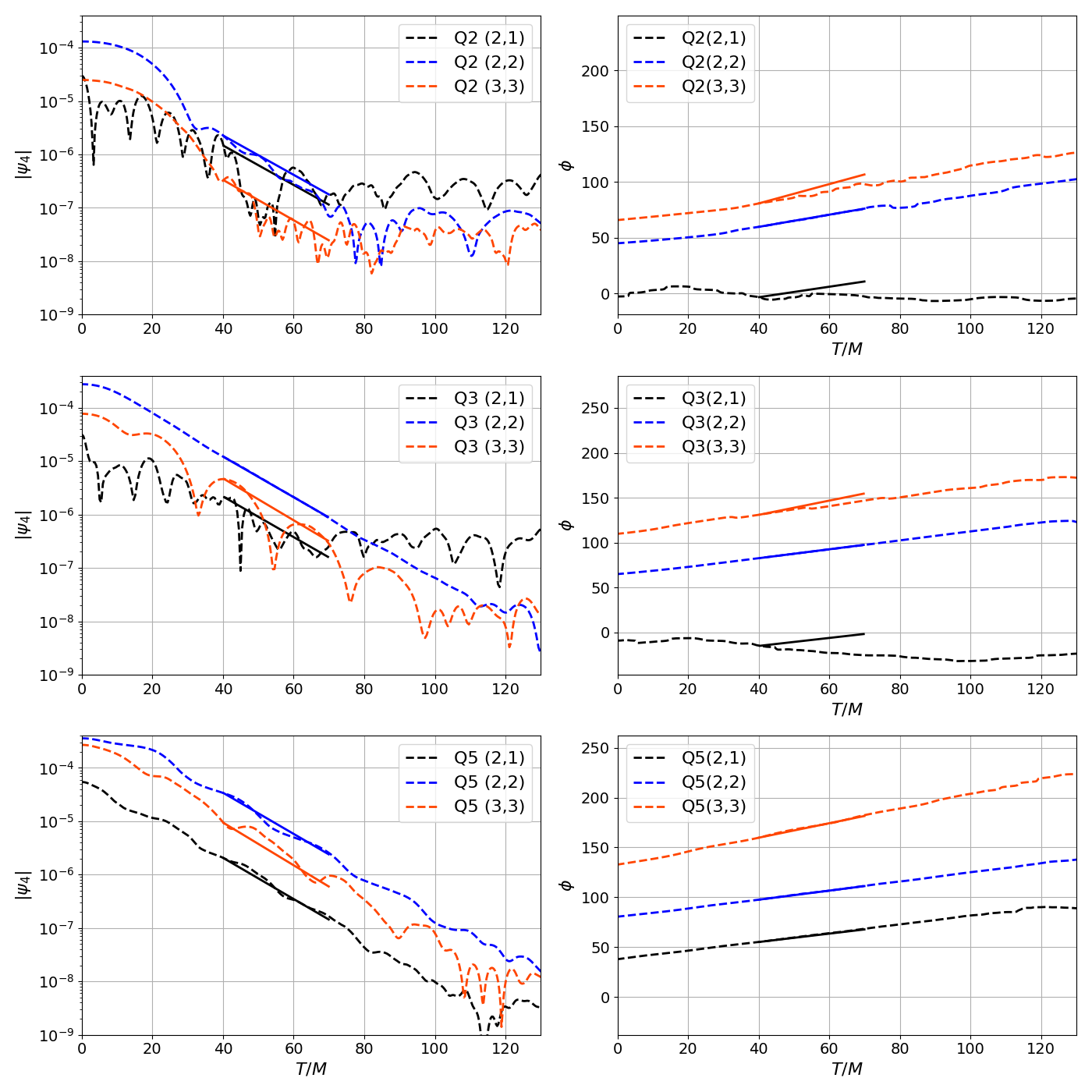}
        \caption{Amplitude (left panels) and phase (right panels) of $\Psi_4$ after peak luminosity for \bhns{} systems. The solid lines are the quasi-normal ringing  computed from Table \ref{tab:qnm_final}.}
\label{fig:qnmfits}
\end{figure}

Next is to discuss the onset of the quasi-normal ringing of the final \bh{}.
Given the mass and the spin of the final \bh{} in Table~\ref{tab:table_remnant}, we compute from the standard fits in the literature~\cite{PhysRevD.73.064030} the quasi-normal frequency and decay time for the (2,1), (2,2) and (3,3) modes. The values are given in Table~\ref{tab:qnm_final}.
Figure~\ref{fig:qnmfits} shows the amplitude (left panels) and phase (right panels) of the (2,1), (2,2) and (3,3) modes of $\Psi_4$ after peak luminosity when the final \bh{} is expected to undergo quasi-normal ringing. In these log-linear for the amplitude and linear-linear for the phase plots, quasi-normal ringing (i.e. exponentially damped sinusoidal) would show up as linear dependence with time for both the amplitude and the phase. For reference, the solid lines are the quasi-normal ringing  computed from Table \ref{tab:qnm_final}.

For the (2,1) mode, we see that the only case showing quasi-normal behaviour is the $q=5$, the one with the more \bbh{-}like characteristics. For the other two cases, there are two factors that prevent a clean quasi-normal ringing. One is that the geometry of the tidal debris does not favor excitation of the final \bh{} in this mode. The other is that, during the time spanned in the figure for the decay of $\Psi_4$ ($\sim 100\,M$), the final \bh{} is still growing as one can see from Figure~\ref{fig:BH_mass}. The (2,2) mode is the one with more noticeable quasi-normal characteristics, in particular in the phase. The  exponential decay of the amplitude is cleaner for the $q=3$ case, and the $q=5$ case shows the bumps associated with the contributions from higher modes due to the center of mass displacement. The $q=2$ case shows exponential decay after $50\,M$, which according to the left panel in Figure~\ref{fig:BH_mass} is when the \bh{} has almost stopped accreting the debris from the disrupted \ns{}. Interestingly, in all cases, the phase of the (3,3) mode  shows an approximate linear growth (i.e. constant frequency of oscillation), but only in the case $q=5$ the growth matches that of quasi-normal ringing. Similarly, exponential decay in the amplitude is not as clear with the exception of the $q=5$. The oscillations in the $q=3$ we conjecture are associated with the accretion of tail of debris observed in the bottom left panel in Figure~\ref{fig:q3_densityplots}.

\begin{table}
\begin{center}
\begin{tabular}{@{}*{10}{l}}
\br
System & $q$ & $\omega_f^{2,1}$  &  $\tau_f^{2,1}$  & $\omega_f^{2,2}$  &  $\tau_f^{2,2}$ & $\omega_f^{3,3}$  &  $\tau_f^{3,3}$  \cr
\mr
\bhns{} & 2 & 0.471 	& 11.73 &  	0.546  &  	11.8  &	0.866 &	11.49 \cr
\bhns{} & 3 & 0.446 	& 11.42	&   0.499  & 	11.42 &	0.794 &	11.06  \cr
\bhns{} & 5 & 0.421 	& 11.26 &  	0.454  &  	11.27 &	0.726 &	10.86 \cr 
\br
\end{tabular}
\caption{Quasi-normal frequencies and damping times computed from the mass and the spin of the final \bh{} in Table~\ref{tab:table_remnant} using the standard fits in the literature~\cite{PhysRevD.73.064030}.}\label{tab:qnm_final}
\end{center}
\end{table}

\begin{table}
\begin{center}
\begin{tabular}{@{}*{8}{l}}
\br
System & $q$ & $\omega^{2,1}$  &  $\tau^{2,1}$  & $\omega^{2,2}$  &  $\tau^{2,2}$ & $\omega^{3,3}$  &  $\tau^{3,3}$ \cr
\mr
\bhns{} & 2 & 0.155   &  55.62   & 0.543 & 9.898 & 0.603 & 9.86     \cr
\bhns{} & 3 & -0.332  &  13.1    & 0.492 & 11.32 & 0.483 & 11.52    \cr
\bhns{} & 5 & 0.436   &  10.83   & 0.449 & 11.08 & 0.71  & 10.08 \cr 
\br
\end{tabular}
\caption{Quasi-normal  frequencies and damping times computed from fitting the data in Figure~\ref{fig:qnmfits}.}\label{tab:qnm}
\end{center}
\end{table}

\begin{figure}
\centering
\includegraphics[width=0.85\textwidth]{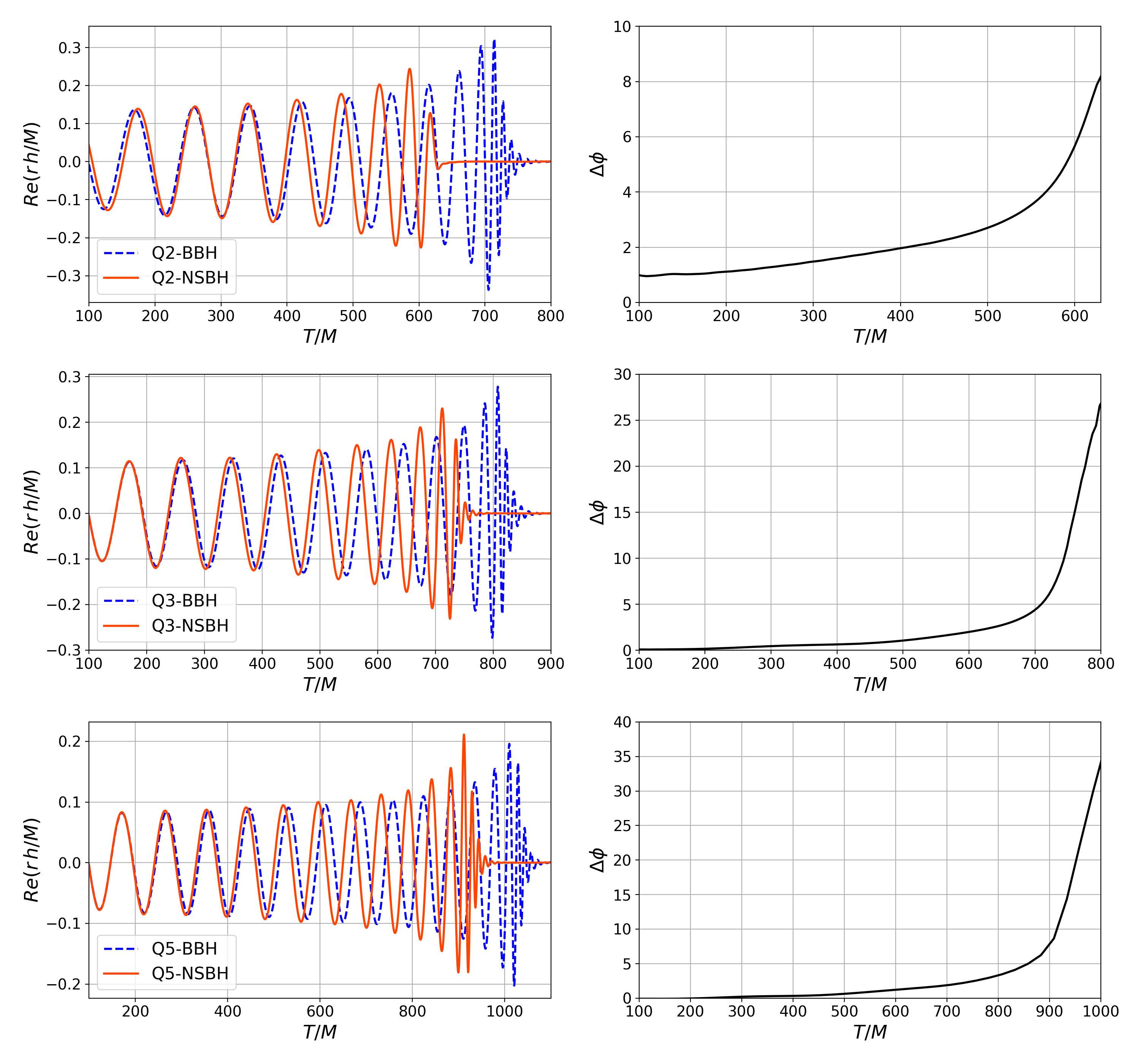}
        \caption{Panels on the left show the real part of  the (2,2) mode of the strain. From top to bottom $q=2,\,3$ and 5. \bhns{} waveforms are depicted with solid lines and the corresponding \bbh{} waveform with dashed lines. Panels on the right show the phase differences between \bbh{} and \bhns{} waveforms.}
\label{fig:ht}
\end{figure}

\begin{figure}
\centering
\includegraphics[width=0.7\textwidth]{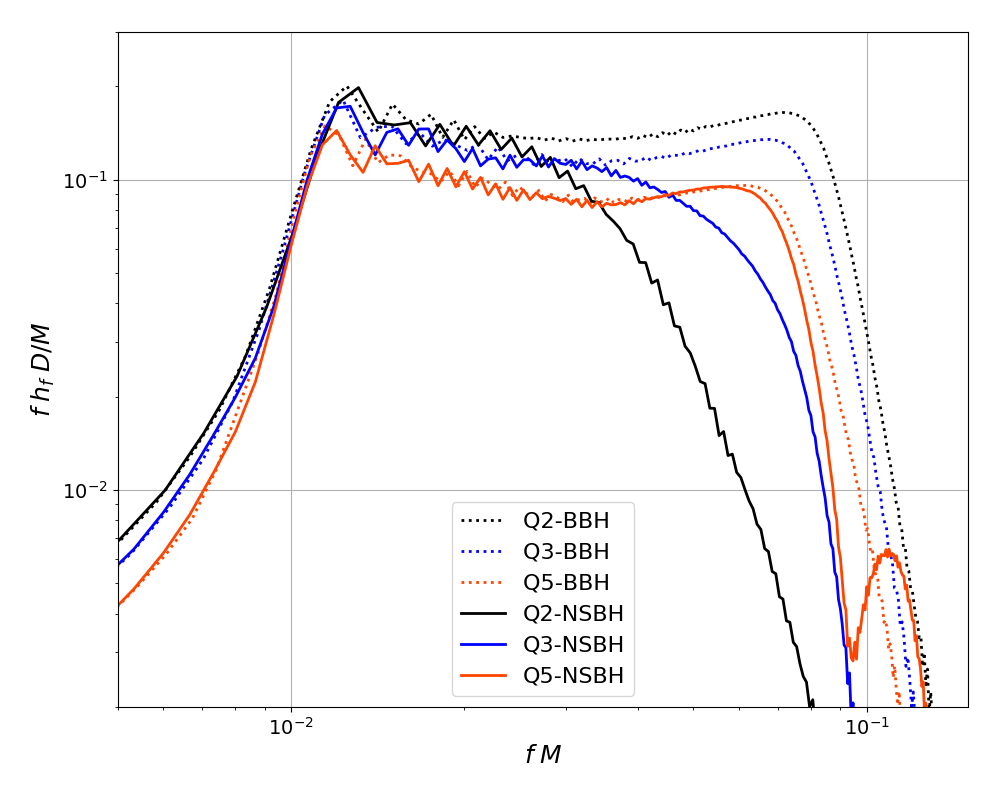}

        \caption[Strain Fourier Spectrum]{Fourier Spectrum of the Strain}
\label{fig:hf}
\end{figure}

\subsection{Spectrum and Mismatches}

Ultimately, comparisons between \bbh{} and \bhns{} systems would be incomplete if not looked through the eyepiece of data analysis tools. The focus of a follow up paper will pay particular attention to observational signatures from the compactness of the \ns{}. For the present work, we start by showing in Figure \ref{fig:ht} the strain of both the \bhns{} and \bbh{} systems for each mass ratio (from top to bottom). Panels on left show the plus polarization of the $(2,2)$ mode of the strain and panels on the right show the phase difference between the \bhns{} and \bbh{} waveforms. We observe very similar characteristics to $\Psi_4$ as seen in Figure \ref{fig:psi4}. The peak of strain in the \bhns{} systems occurs earlier than the \bbh{} for the same mass ratio suggesting early mergers. The two signals overlap for longer durations with increasing mass ratio as seen from the phase differences. The phase differences grow steadily reaching a peak beyond which the \bhns{} signal dies off as the system reaches a stable state. 

Figure \ref{fig:hf} shows the Fourier spectrum of strain of both the \bhns{} and \bbh{} systems. It is evident how the \bhns{} and the corresponding \bbh{} system agree early on for low frequencies, with the $q=5$ following each other through merger.

Next we show in Table~\ref{tab:mismatch} mismatches, $1 - \mathcal{O}(h_1, h_2)$, 
relative to LIGO and the Einstein Telescope, with the matches given by
\begin{equation}
    \mathcal{O}(h_1, h_2) = \frac{\max_{\phi_c, t_c} \langle h_1, h_2 \rangle}{\sqrt{\langle h_1, h_1 \rangle \langle h_2, h_2 \rangle}}
\end{equation}
maximized by the time and phase at coalescence, $t_c$, and $\phi_c$. The inner product $\langle \; , \; \rangle$ is defined as
\begin{equation}
        \langle h_1, h_2 \rangle \equiv 4 Re \int_{0}^{\infty}\frac{\tilde{h}^*_1(f) \tilde{h}_2(f)}{S_n(f)} df
\end{equation}
In this expression, $h$ are strains including modes up to $l=8$ and $S_n(f)$ is the one sided power spectral density of the detector noise. The mismatches in Table~\ref{tab:mismatch} include three different inclinations: $i=0,\,\pi/6$ and $\pi/3$. As expected, for $i=0$ (face-on), the mismatch is highest for $q=2$ because the (2,2) mode dominates in this case; the mismatch decreases with increasing mass ratios consistent with our previous observations. Across the detectors, mismatches are smaller for the Einstein Telescope compared to LIGO, given the higher sensitivity of the former. For $i=\pi/6$, the trend remains similar though the mismatch values decrease for $q=2$ while increase for other two cases because of the contributions from higher modes. This situation becomes more visible for $i=\pi/3$, where the contributions of higher modes increase significantly. The mismatch now increases with mass ratio with $~12\%$ mismatch between BBH and NSBH waveforms for $q=5$ and $~10\%$ mismatch for $q=3$. This shows the importance of higher modes in the study of mixed binaries at high mass ratios.

\begin{table}
\begin{center}
\begin{tabular}{@{}*{5}{l}}
\br
$q$ & Detector  &  $i = 0$ & $i = \pi/6$ & $i = \pi/3$ \cr 
\mr
 2 & LIGO   &  0.0845   & 0.0756 & 0.074  \cr
 2 & ET     &  0.0767   & 0.0686 & 0.0666 \cr
 3 & LIGO   &  0.0611   & 0.0731 & 0.0984 \cr
 3 & ET     &  0.0464   & 0.0576 & 0.0814 \cr
 5 & LIGO   &  0.0046   & 0.0401 & 0.1194 \cr
 5 & ET     &  0.0030   & 0.0349 & 0.1072  \cr 
\br
\end{tabular}
\caption{Mismatches between \bbh{} and \bhns{} waveforms for three different inclination angle $i$ and for the Einstein Telescope (ET) and LIGO.}\label{tab:mismatch}
\end{center}
\end{table}

\section{Conclusions}
\label{sec:conclusions}
For high mass ratio systems, distinguishing \bhns{} binaries from \bbh{} binaries will incur challenges because the \ns{} is swallowed by the hole without experiencing significant disruption. To investigate the transition of the merger behaviour of a \bhns{} into a \bbh{-}like system, we have carried out three \bhns{} merger simulations and their corresponding \bbh{} mergers for mass ratios $q=2,\,3$ and $5$. The \bhns{} system with $q=2$ represents the case of total \ns{} disruption before merger, and the $q=5$ case is an example of a \bbh{-}like merger. The focus was on the effects that the disruption of the \ns{} imprints on the inspiral and merger dynamics, the properties of the final \bh{}, the accretion disk, the \gw{s}, and the strain spectrum and mismatches. A secondary objective of the study was to demonstrate the effectiveness of the method we developed in Ref.~\cite{Clark:2016ppe}
to construct initial data with a generalization of the Bowen-York data for \bh{} punctures to the case of \ns{s}.

The most noticeable feature observed in the simulations of the merger dynamics of the \bhns{} binaries was that they merge earlier than their corresponding \bbh{s}. We found that the dominant factor hardening the mixed binary is the enhanced angular momentum emission carried out by the \gw{s} due to the tidal deformations in the \ns{}. On the other hand, the tidal disruption of the \ns{} suppresses the gravitational recoil of the final \bh{} in \bhns{} mergers when compared with \bbh{s}. Regarding the final \bh{,} its mass is comparable between the \bhns{} and \bbh{} systems. This, however, is not the case regarding the final spin. For instance, in the case of $q=2$, the tidal debris as is accreted by the hole increases the spin by approximately $10\,\%$. The same tidal debris has an influence in the quasi-normal ringing of the final \bh{.} For low $q$'s only the (2,2) mode exhibits a clean damped exponential sinusoidal behavior. In terms of mismatches, the most favorable configuration to distinguish between \bhns{} and \bbh{} systems with large $q$'s would be that for large inclinations where higher modes are more influential.

\paragraph*{Acknowledgements}
This work is supported by NSF grants PHY-1908042, PHY-1806580, PHY-1550461. MGL acknowledges the support from the Mexican National Council of Science and Technology (CONACyT) CVU 391996. We would also like to acknowledge XSEDE  (TG-PHY120016) and the Partnership for an Advanced Computing Environment (PACE) at the Georgia Institute of Technology, Atlanta, Georgia, USA for providing necessary computer resources for this study. We thank Chris Evans, Deborah Ferguson,  Zachariah Etienne and Deirdre Shoemaker for discussions and sharing their resources for this work. We would also like to thank the reviewers for extremely useful feedback and comments. Finally, the authors would also like to express their gratitude to all the front-line workers for their efforts and dedication to keep us all safe during these challenging times.

\vspace{0.25in}
\bibliographystyle{iopart-num}
\bibliography{iopart-num,refs}
\end{document}